\documentclass[twocolumn]{aastex631}
\usepackage{xcolor}
\usepackage[T1]{fontenc}
\usepackage{subfigure}

\graphicspath{{./}{figures/}}
\DeclareUnicodeCharacter{2212}{-}
\begin{document}

\title{Detection  of  Quasi-Periodic Oscillations in the Blazar S4 0954+658 with TESS}

\author{Shubham Kishore}
\affiliation{Aryabhatta Research Institute of observational sciencES (ARIES), Manora Peak, Nainital 263001, India}
\affiliation{Department of Physics, DDU Gorakhpur University, Gorakhpur - 273009, India}

\author{Alok C. Gupta}
\affiliation{Aryabhatta Research Institute of observational sciencES (ARIES), Manora Peak, Nainital 263001, India}
\affiliation{Shanghai Frontiers Science Center of Gravitational Wave Detection, 800 Dongchuan Road, Minhang, Shanghai 200240, People's Republic of China}

\author{Paul J. Wiita}
\affiliation{Department of Physics, The College of New Jersey, 2000 Pennington Rd., Ewing, NJ 08628-0718, USA}

\begin{abstract}
\noindent
 We report the detection of  several quasi-periodicities around 0.6--2.5 days in the optical emission of the blazar S4 0954+658. The source was observed by the Transiting Exoplanet Survey Satellite (TESS) in six sectors and it showed these features in all but one of them, with a QPO of 1.52 days apparently present in portions of four of them.  We used the generalized Lomb-Scargle periodogram method to search for significant signals and we confirmed them using a  weighted wavelet transform for  time-frequency domain analyses. We discuss several possible explanations for these rapid quasi-periodic variations and suggest that an origin in the innermost part of the accretion disk is most likely.  Within this framework, we provide estimates for the mass of the black hole at the core of this blazar.

\end{abstract}

\section{Introduction}\label{sec:intro}
\noindent
Active Galactic Nuclei (AGNs) produce tremendous amount of energy across the entire electromagnetic (EM) spectrum that are not traceable to the stars or their remnants. Blazars belong to the  fraction of AGNs that are radio-loud and that have their relativistic charged particle jets  directed almost along the observer's line of sight ($\leq$ 10$^{\circ}$) \citep{1995PASP..107..803U}. Blazars are usually classified into  two subclasses: BL Lacertae objects (BL Lacs) and flat spectrum radio quasars (FSRQs). BL Lacs show featureless  continua or very weak emission lines (i.e., equivalent width (EW) $\leq$ 5\AA) \citep{1991ApJS...76..813S,1996MNRAS.281..425M}, whereas FSRQs show prominent emission lines in the composite Optical/UV spectra \citep{1978PhyS...17..265B,1997A&A...327...61G}. Blazars manifest large flux and spectral variability in  all observable EM bands and substantial polarization variability from radio to optical bands,  and this emission from blazars is predominantly non-thermal \citep[e.g.,][and references therein]{2019AJ....157...95G}. \\ 
\\The spectral energy distributions (SEDs) of blazars include two broad components (peaks) where the one at lower energy  is dominated by synchrotron emission from the relativistic jet and the  higher energy one is probably due to the inverse Compton (IC) effect \citep[e.g.,][]{2010ApJ...718..279G}. Considering the  recent  nomenclature  applicable to all types of nonthermal dominated AGNs, based on the position of  the lower energy component of the SED, blazars are classified as LSPs (low synchrotron peaked blazars)  if their lower energy component peaks  in the near-infrared (NIR) wavelengths ($\nu_{peak} < 10^{14} $Hz), ISPs (intermediate synchrotron peaked blazars) where the synchrotron emission peaks near the optical energies ($10^{14} $ Hz $ \leq \nu_{peak} \leq 10^{15} $Hz), and HSPs (high synchrotron frequency peaked blazars)  if their lower energy component peaks  in the ultraviolet (UV) or higher energies ($\nu_{peak} \geq 10^{15}$ Hz) \citep{2010ApJ...716...30A}. \\
 \\Quasi-periodic oscillations (QPOs) in  the time series data, or  light curves (LCs), are frequently detected in X-ray binaries (XRBs) \citep{2006ARA&A..44...49R} but rarely discerned in AGNs \citep[e.g.,][and references therein]{2008Natur.455..369G,2014JApA...35..307G,2018Galax...6....1G,2018A&A...616L...6G,2022MNRAS.510.3641R,2022MNRAS.513.5238R}.  Nonetheless, QPOs have occasionally been claimed to have been in various blazars, narrow line Seyfert 1 (NLS1) galaxies, and other subclasses of AGNs in different EM bands on diverse time-scales as short as a few minutes to as long as a few years \citep[e.g,][and references therein]{2008Natur.455..369G,2008ApJ...679..182E,2009ApJ...690..216G,2018A&A...616L...6G,2013ApJ...776L..10L,2014MNRAS.445L..16A,2015MNRAS.449..467A,2014JApA...35..307G,2018Galax...6....1G,2015ApJ...813L..41A,2016ApJ...819L..19P,2016AJ....151...54S,2018A&A...615A.118S,2017ApJ...845...82Z,2017ApJ...849....9Z,2018ApJ...853..193Z,2017ApJ...847....7B,2019MNRAS.487.3990B,2020A&A...642A.129S,2021MNRAS.501...50S,2021MNRAS.501.5997T,2022MNRAS.510.3641R,2022MNRAS.513.5238R}.\\
\\The blazar S4 0954+658 ($\alpha_{2000.0} =$ 09$^{h}$58$^{m}$47.244$^{s}$, $\delta_{2000.0} =$ +65$^{\circ}$33$^{'}$54.8$^{"}$), situated at a redshift  of 0.3694$\pm$0.0011 \citep{2021MNRAS.504.5258B}, is  usually classified as a BL Lac object due to the small equivalent width of the emission lines in its spectrum \citep{1991ApJ...374..431S}; however, this gamma-ray emitter is probably better considered to be a transitional blazar, one falling between the FSRQ and BL Lac categories \citep{2021MNRAS.504.5258B}. This source shows a one-sided radio jet with a polarized hotspot and the polarization of the inner part of the jet indicates a longitudinal magnetic field \citep{1992AJ....104.1687K}. \citet{2011MNRAS.414.2674G} classified this object as an LBL (low energy peaked BL Lac -- same as LSPs) based on the SED. But based on the kinematic features  of the radio jet  \citet{2016A&A...592A..22H} classified this  source as being in their kinematic class II, which is mostly composed of FSRQs; thus S4 0954+658 can well be understood as a transitional object. \\ 
\\Observations with the MAGIC telescopes led to the first detection of S4 0954+658 at  very high energy (VHE) $\gamma$-rays in which   monitoring of the optical polarization angle revealed a rotation of approximately 100$^{\circ}$ \citep{2018A&A...617A..30M}. \citet{2015MNRAS.451L..21B} carried out multi-band optical photometric observations of S4 0954+658 in February 2015 when the source displayed  an unprecedentedly high optical flux state, and found  clear variations on   time-scales of minutes, reaching flux changes  of 0.1 -- 0.2 mag h$^{-1}$. A multi-color photometric and polarization analysis was done by \citet{2015ARep...59..551H} in which the color variability was explained by the superposition of a red component and a bluer variable component with a constant relative SED. Optical variability  detected by \citet{1993A&A...271..344W} showed a large amplitude variation of the order of 100\% on the time-scale of a day. \citet{1999A&A...352...19R} too found large amplitude intra-night variations in optical and radio bands, and that a high optical state  in the SED is associated with a low radio state and vice versa.   They showed that this could be fitted by
a steady-emission helical-jet model that points to the variations being a geometrical effect (involving a changing  jet orientation with respect to the observer’s line of sight). A possible radio-optical correlation was claimed by \cite{1993A&A...271..344W},  but this was based on sparsely sampled radio data during an optical outburst phase. Recently \cite{2021MNRAS.504.5629R} also found a possible 3  week time lag between flux variations in radio and optical bands. \\
\\ \citet{2021MNRAS.504.5629R} carried out  a multi-wavelength observational campaign of S4 0954+658 during 2019 July 18 to 2020 July 18  using the Whole Earth Blazar Telescope (WEBT)\footnote{https://www.oato.inaf.it/blazars/webt/}. Using optical R-band data from WEBT and  early  Transiting Exoplanet Survey Satellite (TESS)\footnote{\href{ http://tess.gsfc.nasa.gov/}{http://tess.gsfc.nasa.gov}} observations for this  source from sectors 14, 20, and 21, they  found a quasi-periodicity of 31.2 days that could be caused by the rotation of an inhomogeneous helical jet whose pitch angle varies in time. \\
\\In 2008, we started a  long-term project to look for QPOs in blazars and NLS1 galaxies in  multiple EM bands on different time-scales. In our 1.5 decades long extensive search, we detected only a few occassional QPOs in blazars \citep[e.g.,][]{2009ApJ...690..216G,2019MNRAS.484.5785G,2009A&A...506L..17L,2009ApJ...696.2170R,2010ApJ...719L.153R,2020MNRAS.499..653K,2020A&A...642A.129S,2021MNRAS.501...50S,2021MNRAS.501.5997T,2022MNRAS.510.3641R,2022MNRAS.513.5238R} and in a NLS1 \citep[e.g.,][]{2018A&A...616L...6G} in different EM bands on  diverse time-scales.  It appears that detectable QPOs in blazars and NLS1 galaxies are quite rare and somewhat difficult to be explained within existing AGN emission models. Here we present a search for optical QPOs in the blazar S4 0954+658 using data taken for the source by TESS  throughout its entire period of operation up to  January 27, 2022.\\
\\The paper is arranged as follows. In Section 2, we  summarize basic information about the TESS satellite and its instrument. In Section 3, we  present the TESS observations of the blazar S4 0954+658 and our  reduction technique for this data. In Section 4, we briefly describe various analysis techniques we have used in search for QPOs. In Section 5,  we give the results of those analyses, indicating the presence of a rather persistent QPO.  A discussion and  conclusions are in Section 6.
\section{Instrument Particulars}\label{sec:Instr}
\noindent
 Employing its four wide-field optical charge-coupled device (CCD) cameras to track at least 200,000 main-sequence dwarf stars  to search for exoplanetary transits \citep{2014SPIE.9143E..20R}, TESS  had a goal of 50 parts per million (ppm) photometric precision on stars with  broad band  \citep{2014SPIE.9143E..20R}  TESS magnitudes 9--15, although it can also  provide extremely useful data on fainter astronomical objects. TESS measures the brightness of preselected target objects every 2 minutes.  As the observations for these targets  are done,  they are read out as ``postage stamps'' and are made available to the community as target pixel files (TPFs) and calibrated LC files. In addition,  TESS records the full frame images  (FFIs) every 30 minutes \citep{2014SPIE.9143E..20R}. These  FFIs  enable users to conduct photometry on any target within the 24$^\circ \times \ \rm{96}^\circ$ field-of-view (FOV).
\begin{table}[h]
\caption{TESS instrument details.}
\hspace*{-1.0cm} 
\resizebox{3.7in}{!}{
\begin{tabular}{c c}\hline\hline
Single camera FOV &  24$^\circ$ $\times$ 24$^\circ$ \\
Combined FOV      &  24$^\circ$ $\times$ 96$^\circ$\\
Entrance pupil diameter & 10.5 cm\\
Focal ratio & f/1.4\\
Wavelength range & 600--1000 nm\\
Ensquared energy & 50\% within 15$\times$15 $\mu$m (1 pix$^2$) \\
                 & 90\% within 60$\times$60 $\mu$m (4 pix$^2$)\\
\hline
\end{tabular}}
\label{tab:inst_det}
\end{table}
\\
\noindent
With a resolution of 21 arcseconds/pixel, the TESS satellite's four cameras each  have a FOV of 24$^\circ \times \ \rm{24}^\circ$ and they are attached in a row, thereby constituting the  total FOV of 24$^\circ \times \ \rm{96}^\circ$.
Initially, the whole sky was divided into 26 strips, called ‘sectors’ (leaving some gaps near the equatorial region), for its sector-wise  supervision. However, subsequent re-observations of  essentially the same portions of  the sky  commenced when the 2 years tenure of  the primary mission was  completed; this increased the number of enumerated sectors.  Because of the rectangular FOV, there are certain regions in the sky that  lie in several sectors (especially near the polar regions). 
  TESS's detailed instrumental specifications are online.\footnote{\href{ https://heasarc.gsfc.nasa.gov/docs/tess/the-tess-space-telescope.html}{https://heasarc.gsfc.nasa.gov/docs/tess/the-tess-space-telescope.html}}
  
\section{Observations and Data Reduction}\label{sec:obs_dat_red}
\begin{figure}[h]
    \vspace{-.5cm}
    \centering
    \resizebox{8.5cm}{!}{\includegraphics{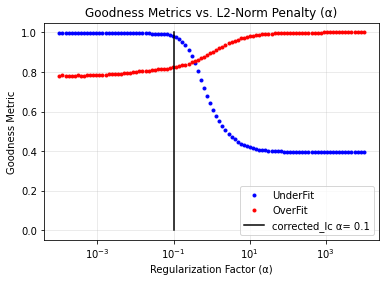}}
    \caption{Goodness metric scan plot for the source 0954+658 observed in sector 14.}
    \label{fig:goodness_metric_plot}
\end{figure}
\vspace{-.4cm}

\noindent
After observing  a sector for roughly 27 days,  TESS creates a targetpixelfile of the preselected source that consists of a series of images taken  at a cadence of 2 minutes. Although an LC can be generated using the targetpixelfile, TESS also creates an LC file that  directly gives a nominal LC of the desired object.  These 2-min cadence TESS LCs are of two types: simple aperture photometry flux (SAP\_FLUX, which is the flux obtained by summing up all the pixel counts in an already defined aperture as a function of time) and pre-search data conditioned simple aperture photometry (PDCSAP\_FLUX),  which is the SAP flux from which long-term trends have been removed);  both are available at Mikulski Archive for Space Telescopes\footnote{\href{https://mast.stsci.edu/portal/Mashup/Clients/Mast/Portal.html}{https://mast.stsci.edu/portal/Mashup/Clients/Mast/Portal.html}}\label{footnote5}. 

\begin{figure*}[t]
    \hspace{-.5cm}
    \resizebox{18cm}{!}{\includegraphics{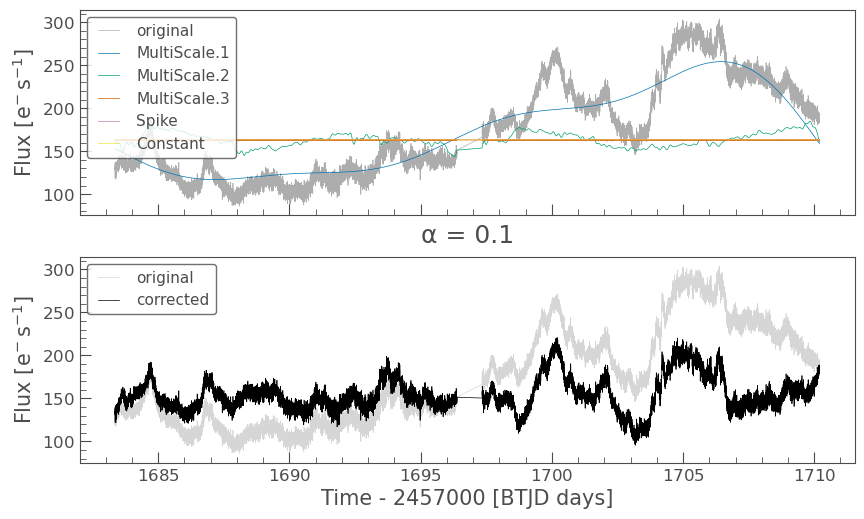}}
    \caption{Diagnostic plot for the source 0954+658 observed in sector 14.  Upper panel:  MultiScale, Spike, and Constant are the different CBVs used for the data reduction.
    Lower panel: the curve labeled original is the raw flux or the SAP\_FLUX, while the Lc after the CBVs and outlier mask are applied is labeled as corrected.
    }
    \label{fig:diag_plot} 
\end{figure*}

\begin{figure*}
    \vspace{-.15cm}
    \hspace{1.5cm}
    \resizebox{15cm}{!}{\includegraphics{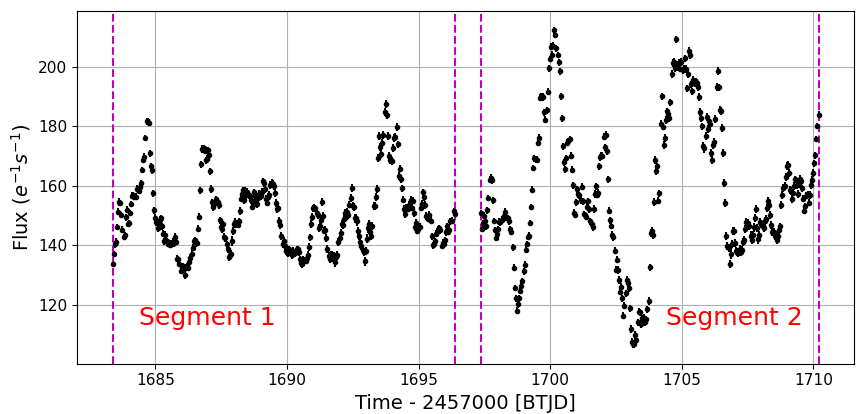}}
    \caption{One hour binned reduced LC of 0954+658 observed in sector 14.\\}
    \label{fig:binned_lc}
\end{figure*}

\vspace*{0.1in}
\noindent
Where the SAP\_FLUX behaves as raw flux data,  a Presearch Data Conditioning (PDC) module used by the TESS science processing operations center (SPOC), gives PDCSAP\_FLUX, after executing a few necessary corrections to the SAP\_FLUX which include the identification and removal of instrumental characteristics introduced due to focus or pointing changes. PDC also accounts for and eliminates isolated outliers \citep{2016SPIE.9913E..3EJ}. In short, the PDCSAP\_FLUX is a sort of refined LC, that is usually cleaner than the SAP flux and has fewer systematic trends;  however, it  may not   exclude instrumental effects producing phony variability which  can be taken care of by subsequent pipelines and it can remove some real longer-term variations present in AGNs.   It is important that TESS also gives the quality of each datum taken  indicated by quality flags so that poor data can be removed. These quality flags, represented by integers, indicate the types of anomalies detected in the data corresponding to each cadence or image. These integers are considered as individual bits that have the meanings described in Table 32 of \citet{twicken2020tess}. The quality flag for a good cadence without any anomalies is set to zero. The quality of each cadence in our analysis has been taken care of in our pipeline, and we used only data points after reduction that have quality flags of zero.\\
 \\
We made use of Lightkurve, a Python package for Kepler and TESS data analysis \citep{2018ascl.soft12013L}, starting our data reduction with  SAP\_FLUX.  We then used co-trending basis vectors (CBVs, a set of systematic trends present in the ensemble of LC data for each CCD) which include three sets of multi-scale basis vectors, a constant vector and spike vectors,  to remove both long term trends and sudden spikes.
The CBVs act like models, the SAP\_FLUX like inputs, and the resultant LC is the output. While reducing the SAP\_FLUX (i.e., fitting the CBVs), there are three parameters that need to be kept in check, viz., underfitting goodness metric, overfitting goodness metric and the regularization parameter ($\alpha$)\footnote{\href{how-to-use-cbvcorrector.html}{http://docs.lightkurve.org/tutorials/2-creating-light-curves/2-3-how-to-use-cbvcorrector.html}}\label{footnote6}. 
 The overfitting metric measures the noise introduced  in the light curve after the correction by measuring the broad-band power spectrum via a Lomb-Scargle Periodogram both before and after the correction. An increase in this power after the correction indicates over-fitting and noise introduction. The underfitting metric measures the mean residual target-to-target Pearson correlation between the target under study and a selection of neighboring targets.  The fitting has been done using L2-Norm/Ridge regression method that includes the third parameter $\alpha$ \citep[the 
ridge regression coefficient estimates are the values that minimize Eq.\ 6.5 of][where $\alpha$ is replaced by $\lambda$]{james2013introduction}. Using the goodness metrics as the Loss Function (difference between data and model) and optimizing the Ridge regularization penalty term, a ridge regression  produces a  set of coefficient estimates for each value of $\alpha$. The goodness metrics are  parts of the {\tt lightkurve.correctors.metrics module}, and we have employed them through the {\tt lightkurve.correctors.CBVCorrector} module.\\
\\
For a good fit, the values of  both the underfitting and overfitting goodness metrics should be kept  at or above 0.8,  and the $\alpha$ should be as low as possible (to make the model fitting a linear regressive).  An example of the values of the two goodness metrics and $\alpha$ we used for one TESS sector of the observations of 0954$+$658 are given in Fig.\ \ref{fig:goodness_metric_plot}.  We have chosen an optimum value,  $\alpha = 0.1$. Corresponding to this $\alpha$, the resulting overfitting and the underfitting metric values found as  results of our calibration pipeline are listed in Table \ref{tab:cal. details} for each sector.  The resultant LCs,  an example of which is shown in Fig.\ \ref{fig:diag_plot}, after removing outliers (data points beyond 3$\sigma$ variance), were then used for the search for any quasi-periodicities.

 \begin{table}[h]
\caption{Flux calibration details}
\hspace*{-1.0cm} 
\resizebox{3.7in}{!}{
\begin{tabular}{c c c}\hline\hline
Sector & Overfitting metric & Underfitting metric\\
\hline
14& 0.823 & 0.982\\
20& 0.913 & 0.993\\
21& 0.833 & 0.991\\
40& 0.969 & 0.999\\
41& 0.967 & 0.998\\
47& 0.908 & 0.976\\
\hline
\end{tabular}}
\label{tab:cal. details}
\end{table}
\noindent

\section{Data Analysis }\label{sec:DATA_anal}
\subsection{Data Segmentation and Binning}
\label{Segmentation_DA}
\noindent
The final LC that is obtained after reduction contains a $\sim$1-day gap in each of the sectors which  arise from one of two reasons. One of these is that the satellite is sending the observed data to Earth during that time span. The gap can also be due to the satellite waiting for any commands to be  uploaded. This gap creates an uneven sampling in the LC, so the  whole LC of one sector is divided into two parts which we call `segments'. After segmenting the LC in each sector, we bin the data in 1 hour chunks  for their further use in the analysis.   Fig.\ \ref{fig:binned_lc} shows this binned LC for Sector 14.

\subsection{Generalized Lomb-Scargle Periodogram (GLSP)}
\label{LSP_DA}
\noindent
The Lomb-Scargle Periodogram (LSP) \citep{1976Ap&SS..39..447L,1982ApJ...263..835S} is a method that is used to detect any periodicity in a time series data (even with irregular sampling) and that uses $\chi ^2$ statistics to fit sine and cosine functions throughout the time series data. The LSP fits a sinusoidal model to the data at each frequency, with a larger power reflecting a highly possible presence of the frequency.
If $\sigma^2$ is the variance, the normalized periodogram or GLSP (\cite{1986ApJ...302..757H}; [Eq. 13.8.4] of \cite{2007nrca.book.....P})
is given as
\begin{equation}
 P(\omega)=\frac{1}{2\sigma^2}\left[\rm{cosine\, term + sine\, term}\right],
 \end{equation}
where cosine and sine terms are given as
\begin{equation}
\rm{cosine\, term}=\frac{\left[\sum_j (X_j-\bar{X})\rm{cos}\, \omega(t_j-\tau)\right]^2}{\sum_j \rm{cos}^2\omega(t_j-\tau)},
\end{equation}
\begin{equation}
\rm{sine}\, term=\frac{\left[\sum_j (X_j-\bar{X})\rm{sin}\, \omega(t_j-\tau)\right]^2}{\sum_j \rm{sin}^2\omega(t_j-\tau)},
\end{equation}
 and where $t_j$ is the time of a measurement, $X_j$ the corresponding flux value, $\bar{X}$ is the mean of $X_j$, $\omega$ is the frequency, 
and $\tau$ is given through 
\begin{equation}
    \rm{tan}(2\omega\tau)=\frac{\sum_j\rm{sin}\,(2\omega t_j)}{\sum_j\rm{cos}\,(2\omega t_j)}.
\end{equation}

\subsection{Wavelet Analysis }
\label{WWZ_DA}
\noindent
Fourier analysis provides an ideal tool to detect a  periodic or quasi-periodic fluctuation, as long as the fluctuation is  present with constant amplitude and constant phase. But astronomical systems do not generally show this constancy and any periodic episode often  appears transiently. Hence  Fourier analysis may not  be optimal where signals  may exhibit short intervals of characteristic oscillation.
As an alternative approach  that can focus its attention on a short and limited time span of the data, the Wavelet Transform  method
is well-suited to detect transient periodic fluctuations.   It can allow for examination of any
 time evolution of the parameters (period, amplitude, phase) describing periodic and quasi-periodic signals. \\
\\A Wavelet is a wave-like oscillation that is localized in time. Wavelets have two basic properties: scale and location. Scale (or dilation) defines how “stretched” or “squished” a wavelet is. Location defines where the wavelet is positioned in time. Following the python code based on a method given  by \citet{1996AJ....112.1709F},
we employed the Weighted Wavelet Z (WWZ)-transform using a Morlet wavelet as a mother wavelet.   This is given as $f(z)=e^{-cz^2}(e^{iz}-e^{-1/4c})$ with $z=\omega(t-\tau)$, where $c$ tells how quickly the wavelet decays and is chosen usually, so that the exponential term decreases significantly in a single cycle  viz. $2\pi/\omega$, where $\omega$ and $\tau$ are  scale factor and  time shift respectively. A proper choice of $c$ for variable lightcurves is less than $1/8\pi^2$ \citep{1996AJ....112.1709F}. We have chosen $c = 0.001$ for our wavelet analysis because such small values of $c$  make the constant $e^{-1/4c}$ essentially negligible and the Morlet wavelet reduces to
\begin{equation}
    f(z)=e^{-cz^2+iz}.
\end{equation} 
Then the WWZ map that is created using  convolution of the LC and the wavelet (scale factor and time-shift dependent) containing the data decomposed in time and frequency domain is given as 
\begin{equation}
    W[\omega, \tau ; x(t_\alpha)]= \omega^{1/2}\sum_{\alpha=1}^{N} x(t_\alpha)f^*[\omega(t_\alpha − \tau )] .
    \label{DWT_form}
\end{equation}
Eq.\ (\ref{DWT_form})is essentially similar to a windowed Fourier transform, with window $e^{-c\omega^2(t-\tau)^2}$. The end result is a two-dimensional plot showing amount of power as a function of both period (or frequency) and time.  Detailed information on the  WWZ can be found in \cite{1996AJ....112.1709F} and \cite{2004JAVSO..32...41T}.

\subsection{LC simulation and Peak Significance}
\label{SIM_DA}
\noindent
To inspect the significance of any peculiar peak obtained in the GLSP and WWZ of LC in  any particular segment, simulations using Monte Carlo method were done  to create artificial LCs
 using the parameters calculated after fitting the Power Spectral Density (PSD) of the  same segment's LC. We followed the method given in \cite{2013MNRAS.433..907E}  in which firstly, the PSD of the segment's LC was fitted with a  power-law distribution  $P(\nu)=A\nu^{-\alpha}$ where $P$ is the power at  frequency $\nu$ with  normalization constant $A$ and spectral index $\alpha$, and using these two parameters ($A\  \&\  \alpha$) obtained from the fit, 5000  individual LCs were generated  using a Monte Carlo simulation for each segment.  The power law fit values are listed in Table \ref{tab:PSD parameters}.
  The spectral indices of the source in different segments here lie in the range between 1.43 and 2.41  having a mean of $1.96\pm0.28$ that is quite typical for the optical variability of blazars \citep{2020ApJ...903..134C}. 
  The normalization constants, however, have a somewhat wider range. A GLSP for each of these pseudo-LCs was then obtained individually. With that large number of GLSPs at each frequency, the mean GLSP with its standard deviation was used to estimate the  significance of any peaks.

 \begin{table}[h]
\caption{PSD fit parameters for each segment}
\hspace*{-1.7cm} 
\resizebox{4in}{!}{
\begin{tabular}{c c c c}\hline\hline
Sector & Segment & Spectral   & Normalization\\
       &         & index ($\alpha$)& constant ($A$)\\    
\hline
14& 1 & 2.13 & 1.1$\times 10^{-5}$ \\
  & 2 & 2.12 & 2.4$\times 10^{-5}$ \\
20& 1 & 1.94 & 2.3$\times 10^{-5}$ \\
  & 2 & 2.15 & 1.4$\times 10^{-5}$ \\
21& 1 & 1.71 & 2.5$\times 10^{-5}$ \\
  & 2 & 1.43 & 2.2$\times 10^{-4}$ \\
40& 1 & 1.71 & 2.0$\times 10^{-5}$ \\
  & 2 & 2.00 & 1.2$\times 10^{-5}$ \\
41& 1 & 1.90 & 1.1$\times 10^{-5}$ \\
  & 2 & 1.67 & 3.5$\times 10^{-5}$ \\
47& 1 & 2.37 & 1.5$\times 10^{-5}$ \\
  & 2 & 2.41 & 0.8$\times 10^{-5}$ \\
\hline
\end{tabular}}
\label{tab:PSD parameters}
\end{table}

\section{RESULTS}
\label{sec:Result}
\begin{table*}[t]
\caption{Segment-wise significant peaks, corresponding periods and their significances}

\resizebox{6.7in}{!}{
    \begin{tabular}{c c c c c c} \hline \hline
    Sector & Segment & Peak Frequency  & Period  &  LSP Significance & TAP significance \\
           &         &  (day$^{-1}$)   & (days)  &  level ($\sigma$) &  level ($\sigma$)\\\hline
     14  &  1  &  0.43  &  2.33  &  8.6  &  12.0\\
          &    &  0.74  &  1.35  &  11.6  &  11.6\\
        &  2  &  0.20  &  5.00  &  4.2  &  11.6\\
     20  &  1  &  0.65  &  1.54  &  10.0  &  11.3\\
      &    &  0.90  &  1.11  &  5.7  &  5.0\\  
      &  2 &  0.36  &  2.78  &  4.5  &  6.3\\
      &    &  0.61  &  1.64  &  6.1  &  6.4\\
      &    &  0.73  &  1.37  &  9.4  &  9.2\\
      &    &  0.92  &  1.09  &  5.8  &  5.3\\
    21&  1  &  --  &  --  &  --  &  --\\
      &  2  &  --  &  --  &  --  &  --\\
  40  &  1  &  0.33  &  3.03  &  6.8  &  12.1\\  
      &  &  0.65  &  1.54  &  6.1  &  5.7\\
        &  2  &  0.72  &  1.39  &  5.7  &  4.1\\
      &    &  0.83  &  1.20  &  10.7  &  7.6\\
  41  &  1  &  0.56  &  1.79  &  6.3  &  8.0\\
       &  2  &  0.66  &  1.52  &  10.3  &  13.3\\
  47  &  1  &  0.23  &  4.35  &  4.6  &  9.2\\
      &    &  0.80  &  1.25  &  8.6  &  6.5\\
      &  2  &  0.50  &  2.00  &  10.9  &  12.7\\
      &    &  0.66  &  1.52  &    17.5  &  18.9\\
      &    &  1.00  &  1.00  &  13.0  &    10.4\\
      &    &  1.13  &    0.88  &  16.6  &  12.9\\
      &    &  1.49  &  0.67  &  22.2  &  12.6\\
      &    &  1.69  &  0.59  &  26.6  &  13.5\\

\hline
\end{tabular}}
\label{tab:Reult}
\end{table*}

\begin{figure*}
    \hspace{.5cm}
    \vspace{-.5cm}
    \resizebox{17cm}{!}{\includegraphics{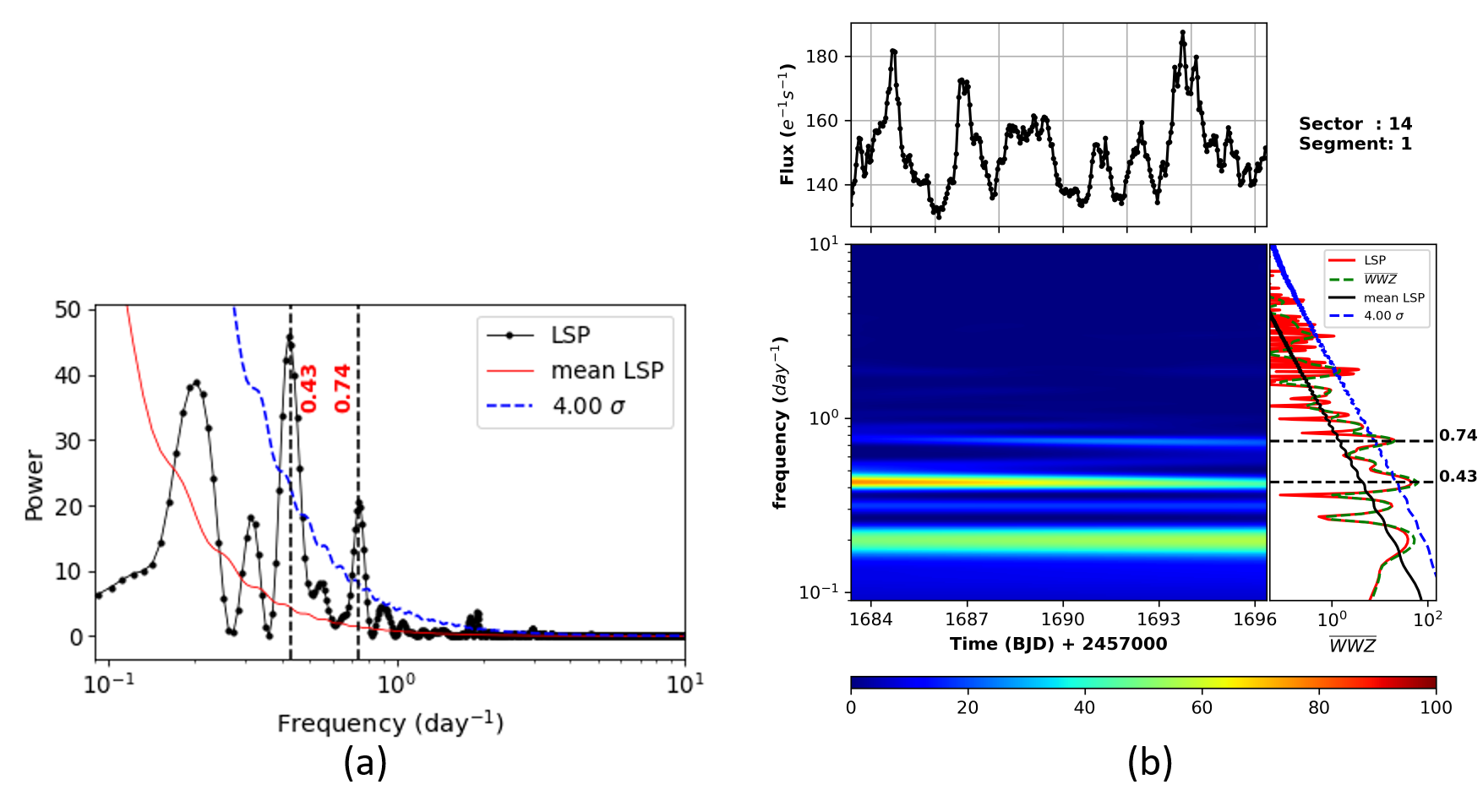}}
    \caption{(a) The GLSP of Sector 14 Segment 1 LC. (b) The upper plot is the 1 hour binned LC corresponding to the same segment.  The lower left figure is the WWZ transform of this LC, showing the power in time and frequency domains. The lower right panel includes the time-averaged periodogram of the WWZ transform, the GLSP, the mean LSP (GLSP-fit) and the $4\sigma$ significance level. 
    }
    \label{fig:14.1}
\end{figure*}

\noindent
In this  work, we have collected and analyzed the data for S4 0954+658 from  TESS, which  observed it in a total of 6 sectors (14, 20, 21, 40, 41 \& 47)  until now.  We  divided them into a total of 12 segments because as described in  Sect.\ \ref{Segmentation_DA}], a gap of roughly one day was  present at the center of each sector.
We present here the results of  our analyses of these segments, and start with the first one, Sector 14, Segment 1.
To detect any periodicity, we  first used the GLSP   and to  check the results of  our GLSP analysis, we then performed the WWZ analysis  that gives the power as a function of time and frequency.  From this WWZ map,  a time averaged periodogram  (TAP), or mean of the WWZ, was obtained. The GLSP and the TAP for this segment have been plotted simulataneously in  the lower right panel of Fig.\ \ref{fig:14.1}b to allow for a comparison of their normalizations  and peaks.   \\
\\To find the significance level, we followed the method described in Sect.\ \ref{SIM_DA} and found different  possible QPOs at various significance levels for different Segments.  We quote significance levels  of any particular frequency only if the GLSP peak and TAP peak corresponding to that frequency both touch or cross the  4$\sigma$ level.  We provide the results of these analyses  in Table\ \ref{tab:Reult}  and   discuss each segment briefly below.

\subsection{Sector 14}
\paragraph{Segment 1}
Visual inspection of  the LC obtained in this segment provides an indication of a probable  QPO. The GLSP (Fig.\ \ref{fig:14.1}a) of this LC shows peaks   at multiple  frequencies:  0.20 day$^{-1}$, 0.43 day$^{-1}$ and 0.74 day$^{-1}$. \\
The peak at 0.43 day$^{-1}$ touches the  $8.6\sigma$  and the one at 0.74 day$^{-1}$  touches the $11.6\sigma$ level. These values are confirmed by the TAP obtained from the WWZ (Fig.\ \ref{fig:14.1}b) method. The corresponding QPO periods are  2.33 days and 1.35 days. The WWZ plot indicates that the 0.43 day$^{-1}$ signal weakens somewhat during the course of this observation  but remains clearly visible throughout the time span, while the one at 0.74 day$^{-1}$  undergoes a gradual rise in strength, followed by a decline, and is most visible for $\sim$7 days corresponding to over 5 cycles during this segment.  Although there is a good amount of power also  present at 0.20 day$^{-1}$,   it  barely touches the 1.5$\sigma$ level. In any event, such frequencies are too low to consider further since the corresponding 5 days period would show fewer than 3 cycles  which could easily arise by chance from red-noise \citep{1978ComAp...7..103P}. When the GLSP is looked upon closely at higher frequencies near 1.8 day$^{-1}$, we see evidence  for two peaks  crossing the  4$\sigma$ level; however, these putative QPOs have  very low powers compared to the noted ones.\\
\paragraph{Segment 2}
  The LC of this segment, along with those of other segments considered in this paper are displayed in Fig.\ \ref{fig:Sector-wise LCs}.  The GLSP and the TAP for the LC of this segment,  shown in Fig.\ \ref{fig:Picture5}A, both peak at 0.2 day$^{-1}$. 
  There is a  difference in the powers obtained by the two methods, with the GLSP being the weaker one, though both are at  or above the 4$\sigma$ level.  The peak has a corresponding period of 5 days spanning the whole segment which, however, as noted above is too long to be taken seriously within a single segment.  But given that there was also an indication of such a period in the immediately preceding Segment 1, we can  say that there may be a signal there that is present for  5 cycles. Considering this, an additional analysis was performed to inspect this entire sector,  but we did not find the frequency peak corresponding to the 5 day period beyond the 4 sigma level.

 \subsection{Sector 20}
 \paragraph{Segment 1}
 The LC shows a great deal of variability that could indicate a possible QPO.  Both the GLSP power and the WWZ plot and TAP power shown in Fig.\ \ref{fig:Picture5}B  agrees with that assessment and has a peak at 0.65 day$^{-1}$  covering the whole segment except near the very end.  Its LSP and TAP peaks have  significance levels of  10.0$\sigma$ and 11.3$\sigma$, respectively,
 corresponding to a period of 1.54 days, so the  periodicity spans over 8 cycles.  The apparent peak at lower frequencies is not significant.   A peak with  substantial power is also present at 0.90 day$^{-1}$, and can be traced  throughout the segment, corresponding to the period of 1.11 days.
 \paragraph{Segment 2}
  There are many oscillations in this LC shown  that might suggest the presence of a QPO in this segment as well.  The results of GLSP and TAP from WWZ shown in Fig.\ \ref{fig:Picture6}C yield various peaks in its periodogram at frequencies: 0.36 days$^{-1}$, 0.61 days$^{-1}$, 0.73 days$^{-1}$ and 0.92 days$^{-1}$, all of which reach a nominal significance level of 4.5$\sigma$. The corresponding periodicities  are: 2.78 days, 1.64 days, 1.37 days and 1.09 days. However, in that, the strongest signal is at the lowest of those frequencies   corresponds to fewer than 5 cycles, while the WWZ plot shows it weakening during the course of this observation, we cannot claim a firm detection.  The next two  higher frequency putative signals bracket the one seen at 0.65 day$^{-1}$ in Segment 1  of the same sector. These two peaks have a mild presence during the whole segment involving a gradual rise near at the end. In addition, there is another at 0.92 day$^{-1}$ with its  weak presence in the second half of the span  for $\sim$8  days.     

 \subsection{Sector 21}
\begin{figure}[h]
    \centering
    \hspace{-1.25cm}
    \resizebox{19.25cm}{!}{\includegraphics{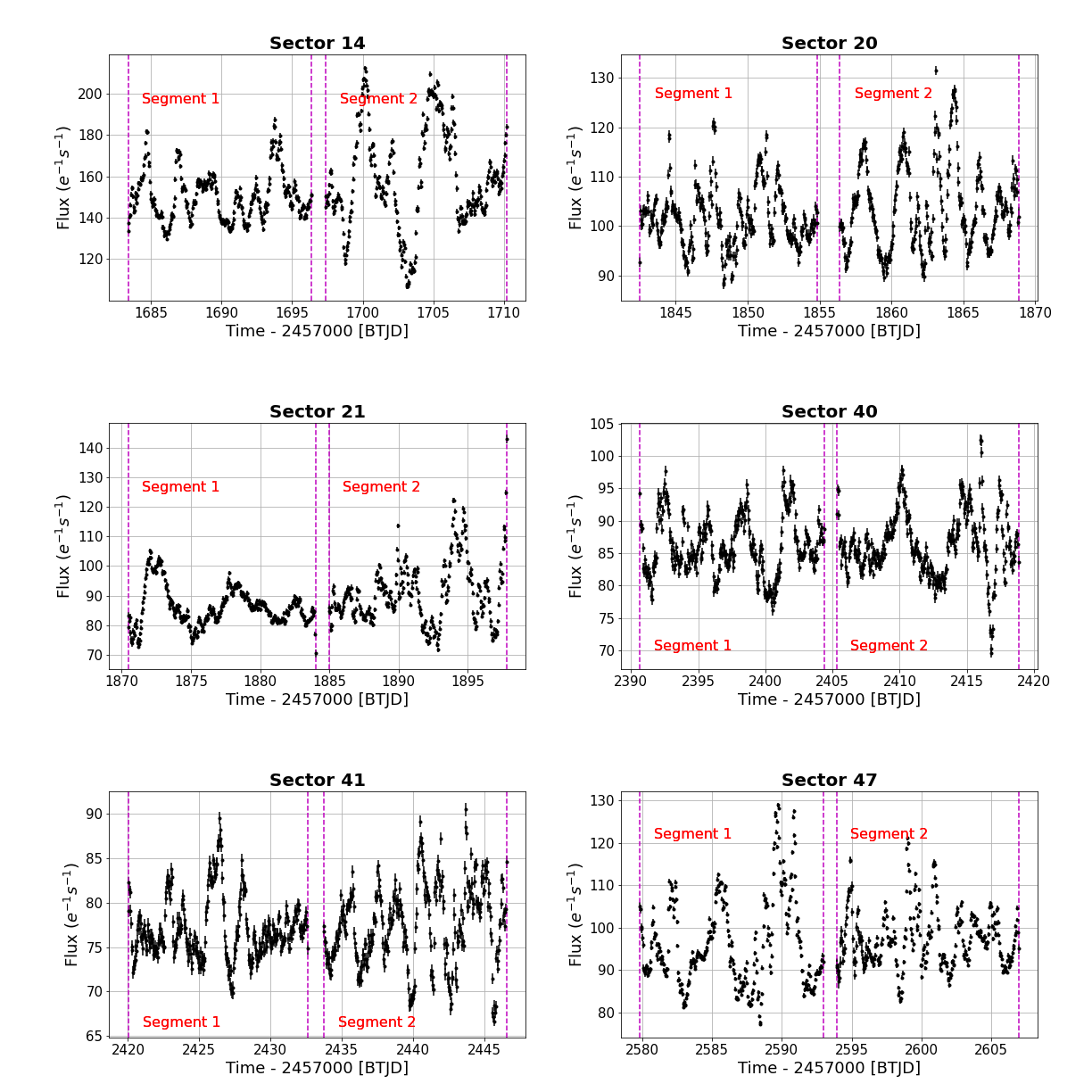}}
    \caption{1 hour binned LCs of S4 0954+658 in all observed sectors.  }

    \label{fig:Sector-wise LCs}
\end{figure}

\begin{figure}[h]
    \centering
    \resizebox{18.5cm}{!}{\includegraphics{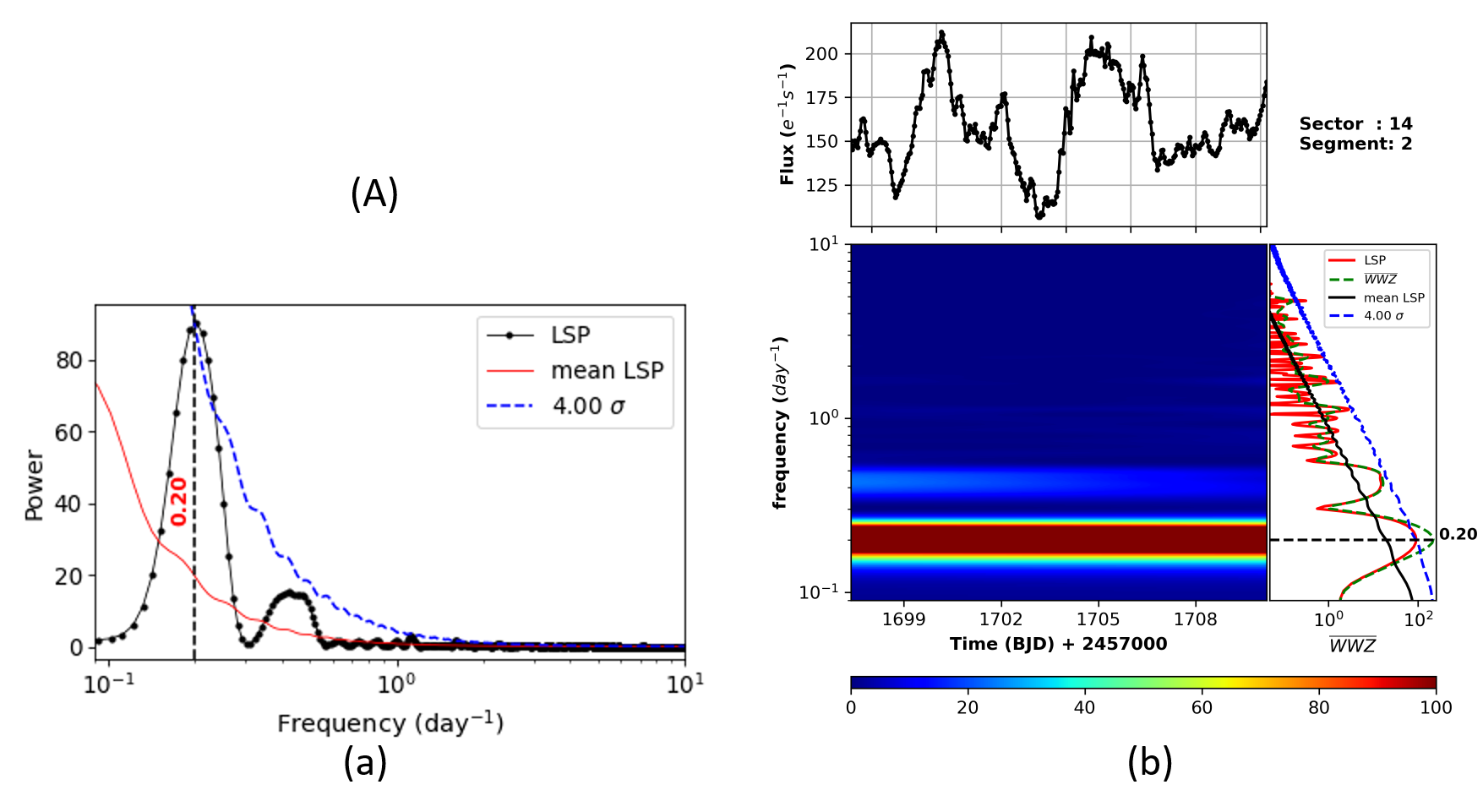}}
    \vspace{1cm}
    \resizebox{18.5cm}{!}{\includegraphics{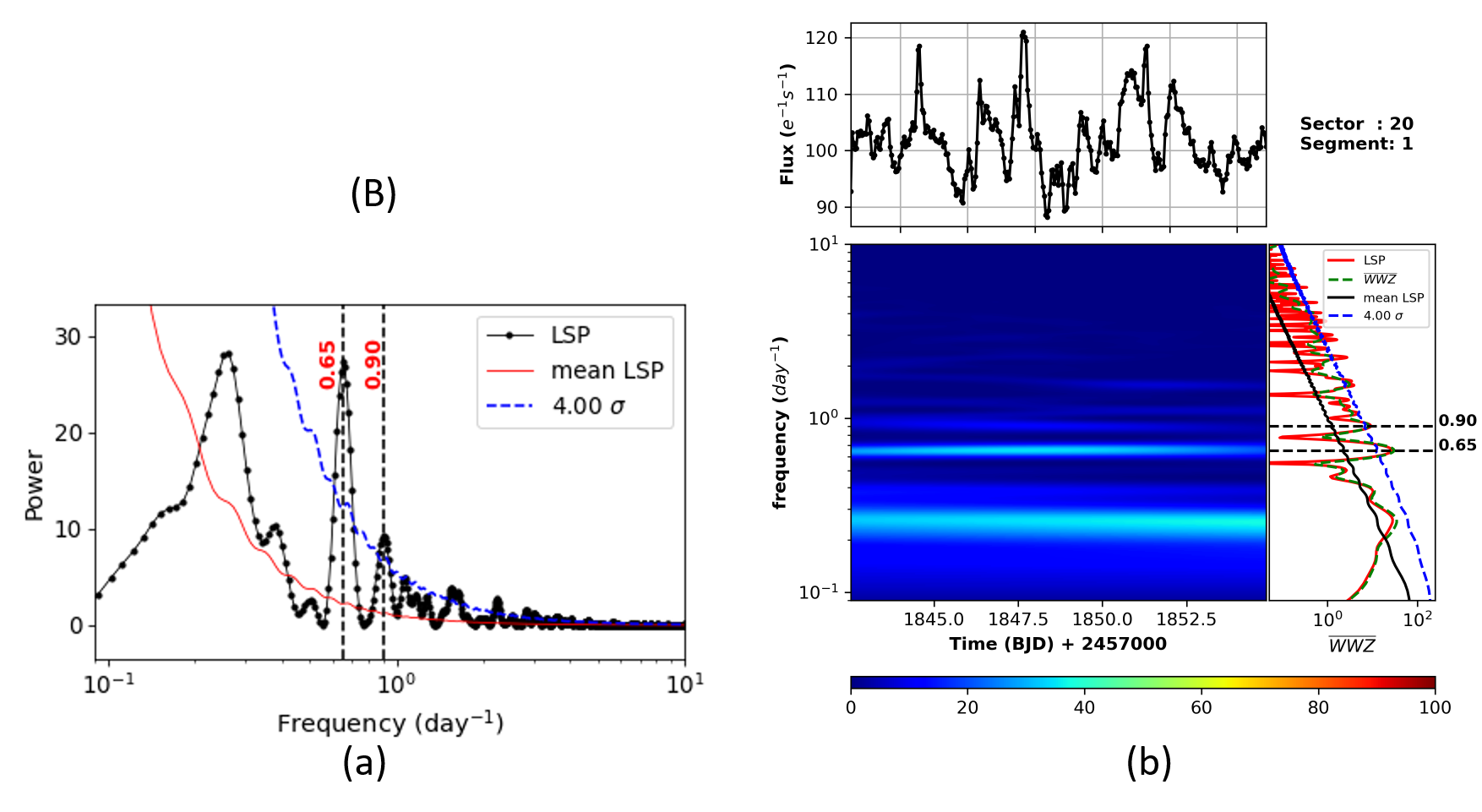}}
    \caption{The upper-case letters label the GLSP and WWZ plots for individual segments in each sector, here the second segment for Sector 14 and first for Sector 20. The lower-case letters within each panel denote the same GLSP, binned LC, WWZ, mean powers and 4$\sigma$ significance level as in Fig.\ \ref{fig:14.1}.}
    \label{fig:Picture5}
\end{figure}

\begin{figure}[h]
    \centering
    \resizebox{19cm}{!}{\includegraphics{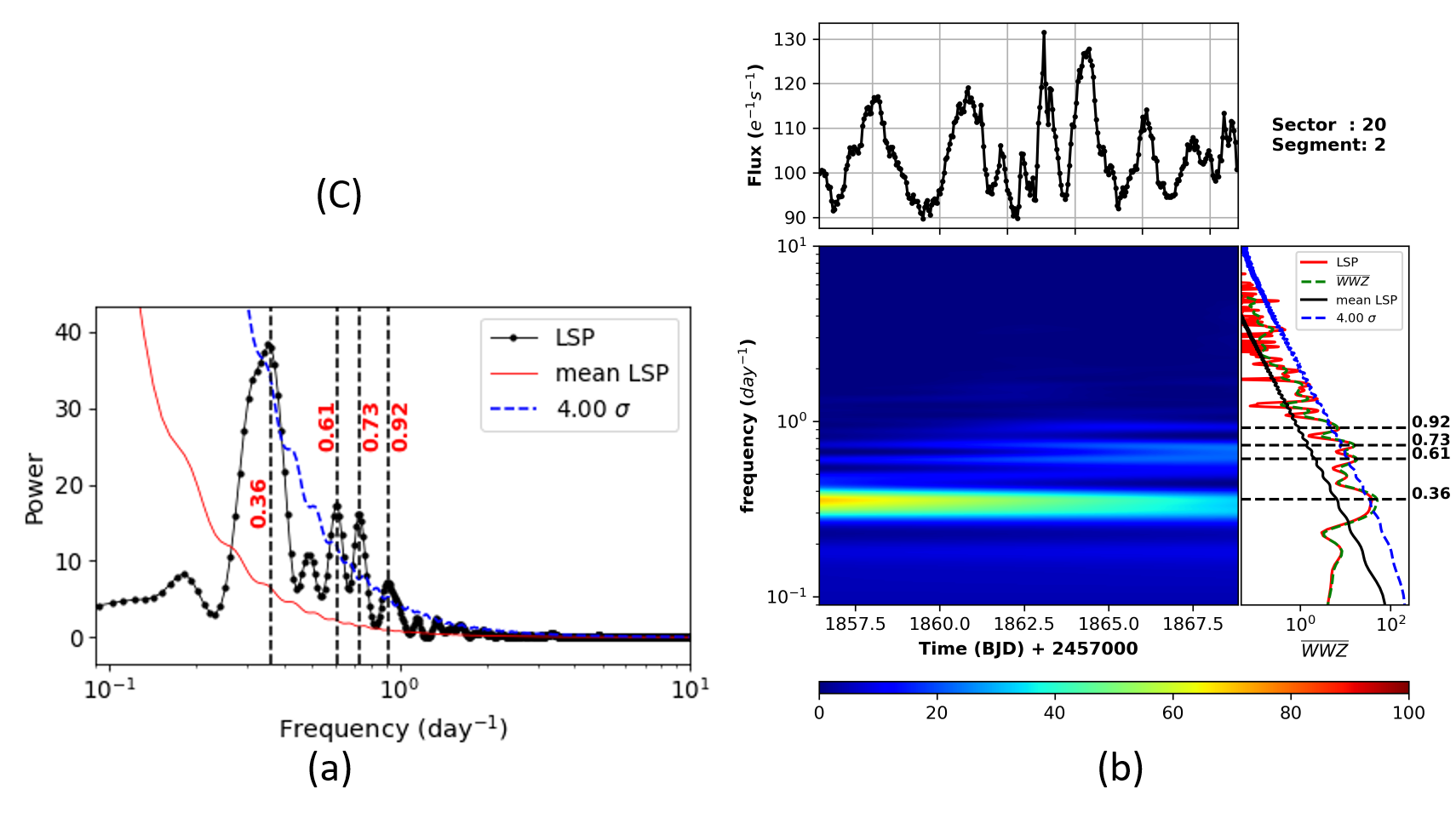}}
    \vspace{1cm}
    \resizebox{19cm}{!}{\includegraphics{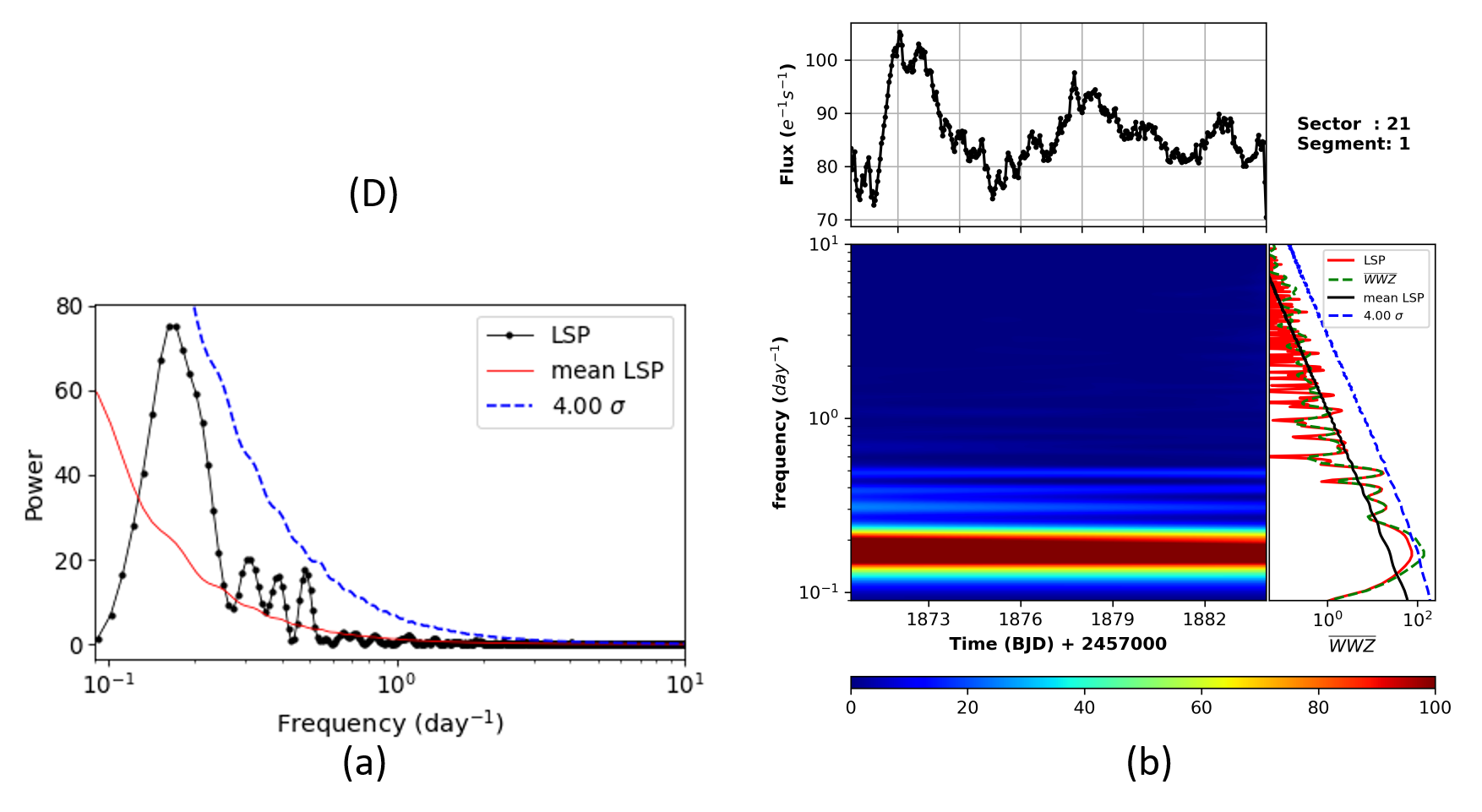}}
    \label{fig:Picture6}
\end{figure}

  The LCs in both segments in Fig.\ \ref{fig:Sector-wise LCs} are definitely variable, but neither hints at a QPO.  The GLSP and WWZ plots shown in Figs.\ (\ref{fig:Picture6}D,\ \ref{fig:Picture7}E) for Segments 1 and 2, respectively,  give no evidence for a signal that  could go beyond a 4$\sigma$ significance level, and we conclude that there was none at this epoch.

 \subsection{Sector 40}
\begin{figure}[h]
    \centering
    \resizebox{19cm}{!}{\includegraphics{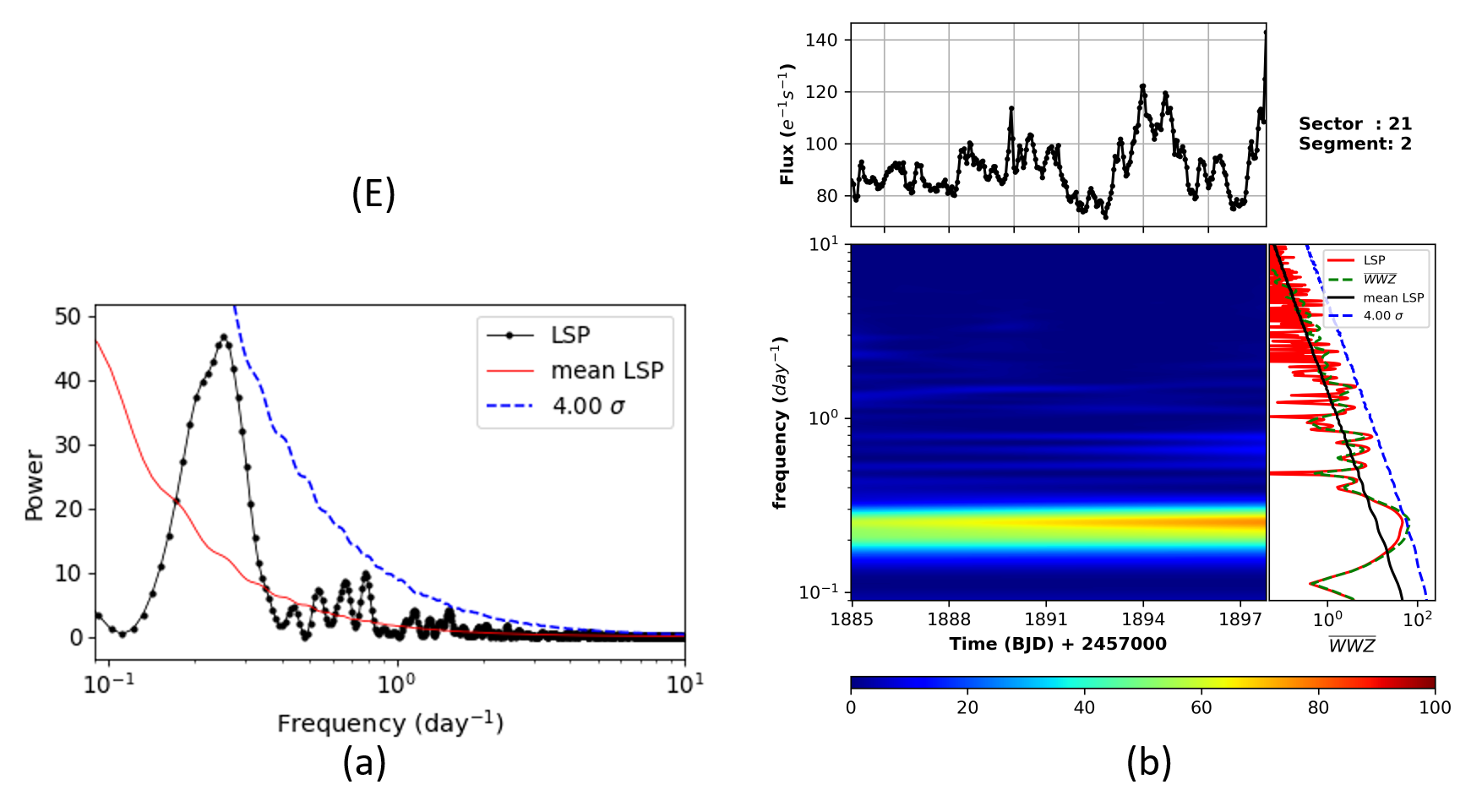}}
    \vspace{1cm}
    \resizebox{19cm}{!}{\includegraphics{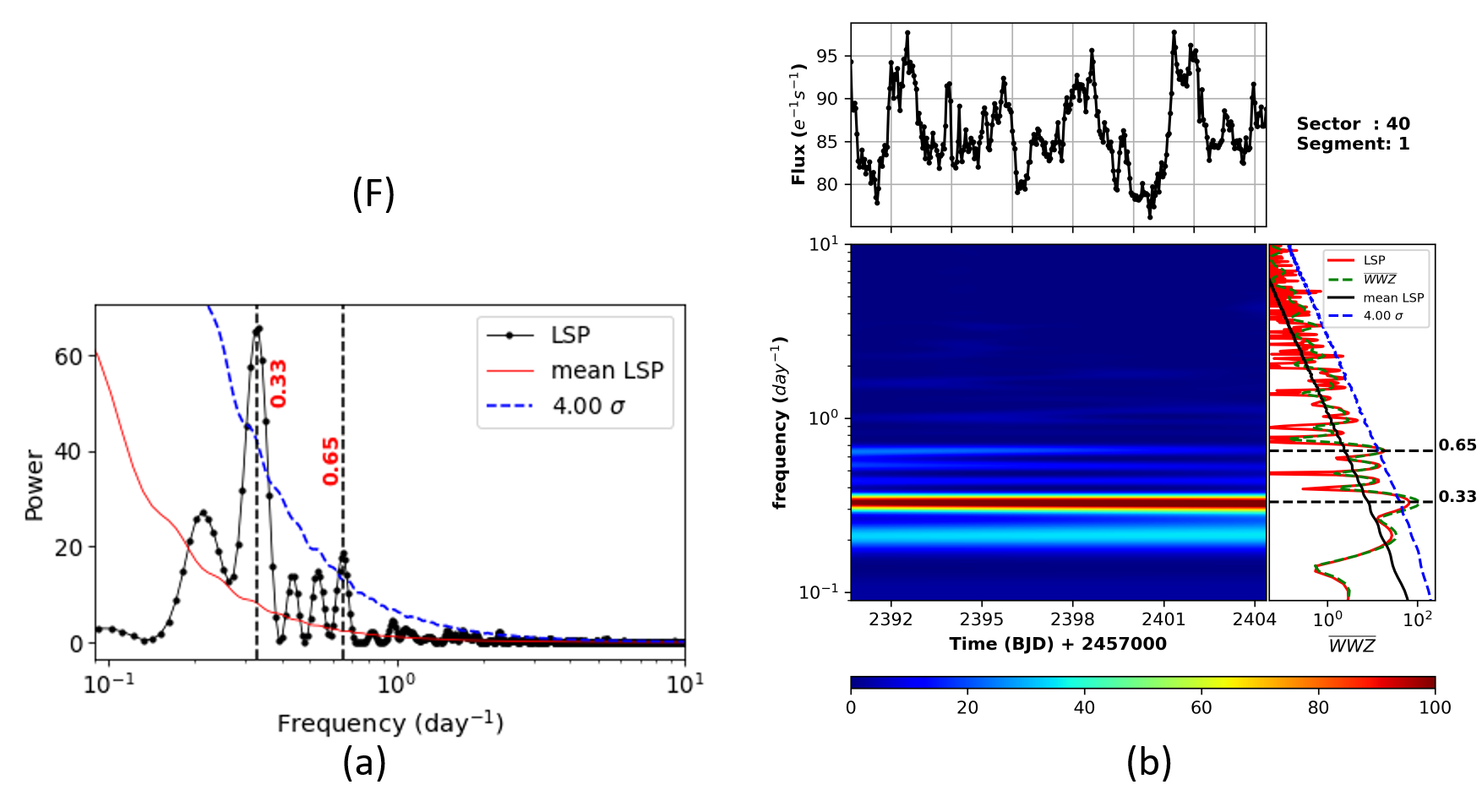}}
    \label{fig:Picture7}
\end{figure}

 \paragraph{Segment 1}
  The GLSP and TAP powers in Fig.\ \ref{fig:Picture7}F show peaks providing evidence of  a QPO at frequency 0.33 day$^{-1}$ present throughout the segment at 6.8$\sigma$ and 12.1$\sigma$ levels with its apparent harmonic at 0.65 day$^{-1}$  also showing up quite strongly. The corresponding periods are 3.03 days and 1.54 days, so again we cannot give full credence to the former as it would encompass only 4 cycles.  However, the latter is more trustworthy and is interestingly close to the signals  present in both segments of Sector 20,  here with power decaying toward the end of the segment.
 \paragraph{Segment 2}
 The GLSP and the TAP  obtained from the WWZ have peaks  shown in Fig.\ \ref{fig:Picture8}G at    0.72 day$^{-1}$ at 5.7$\sigma$ and 4.1$\sigma$ levels, or 1.39 days, and 0.83 day$^{-1}$ at 10.7$\sigma$ and 7.6$\sigma$ levels with a corresponding period of 1.20 days.  However, these peaks have lower power, but higher significance,  compared to    another peak  at  $\sim$0.20 day$^{-1}$, but its low nominal significance and such a long period imply that it cannot be taken seriously. Still, it is worth noting that this peak is also apparent in Segment 1 of the same sector.
 
 \subsection{Sector 41}
\begin{figure}[h]
    \centering
    \resizebox{19cm}{!}{\includegraphics{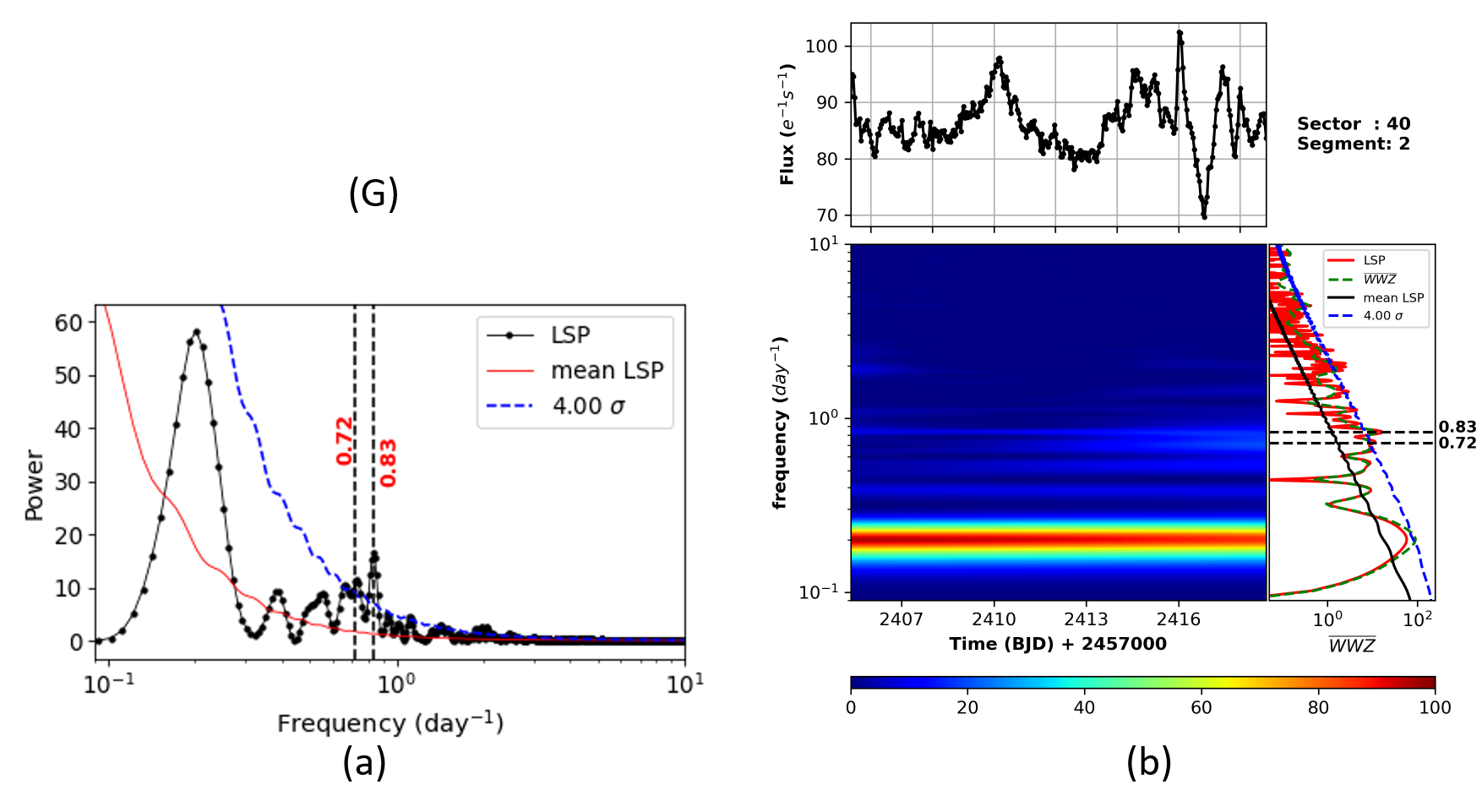}}
    \vspace{1cm}
    \resizebox{19cm}{!}{\includegraphics{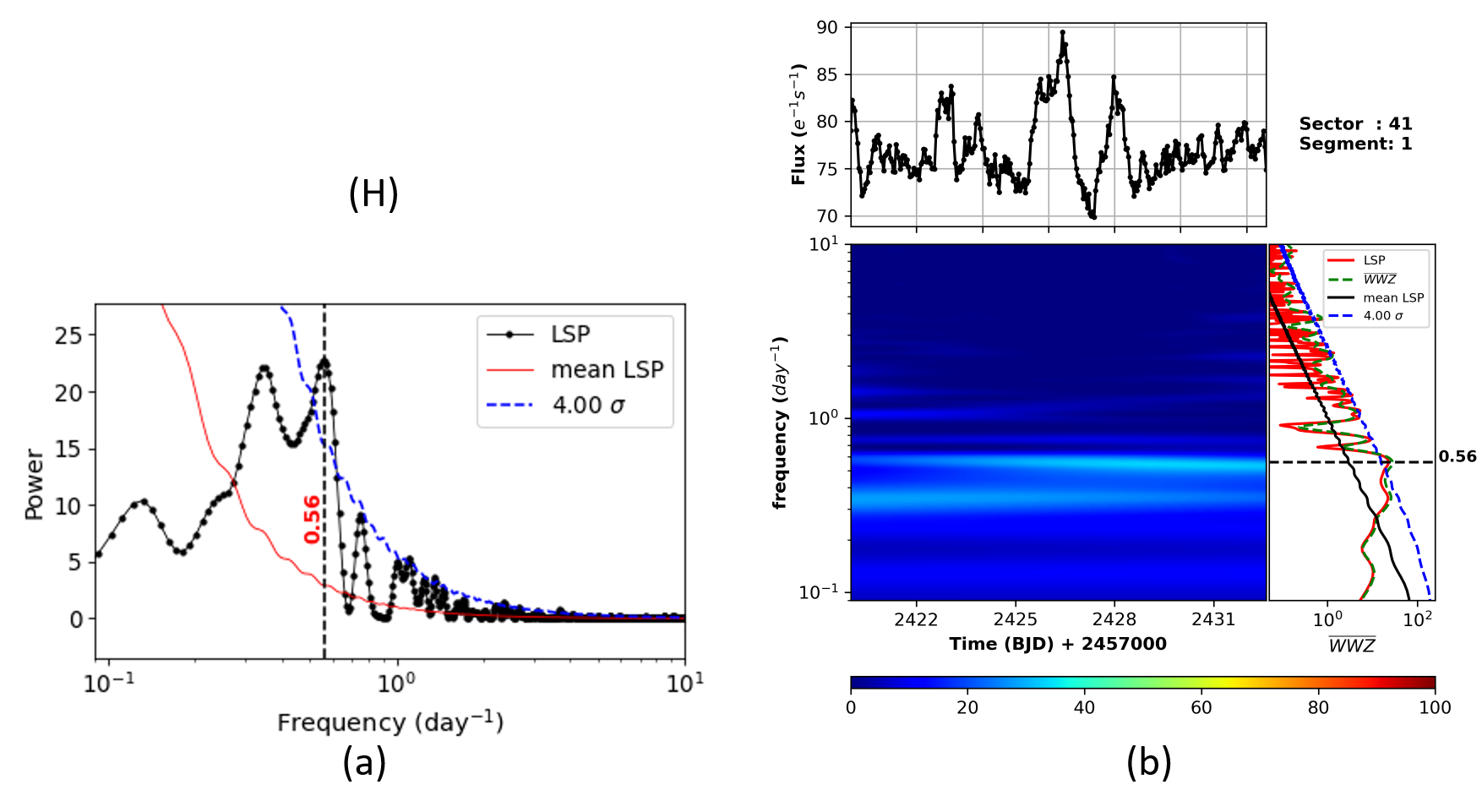}}
    \label{fig:Picture8}
\end{figure}

 \paragraph{Segment 1}
  The LC shows significant fluctuations with a hint of periodicity. In Fig.\ \ref{fig:Picture8}H, showing the  GLSP, WWZ and its TAP, we observe a smeared signal centered at 0.56 d$^{-1}$, with a corresponding period of 1.79 days at  6.3$\sigma$ and 8.0$\sigma$ significance levels in GLSP and TAP, respectively. The WWZ plot indicates that this QPO gains strength after $\sim2$ days from the beginning of this epoch and is present until its end, so for over 6 cycles.  Another peak with a similar power level appears at $\sim$0.35 day$^{-1}$ but  with a lower significance. This signal, however, is mostly present in the former part of the time domain, and then gradually fades.
 \paragraph{Segment 2}
  In Fig.\ \ref{fig:Picture9}I we see that there is a signal at the same frequency seen in two previous Segments, 0.66 day$^{-1}$ with period 1.52 days, reaching a significance levels of 10.3$\sigma$  and 13.3$\sigma$ respectively, in its GLSP and TAP.  This peak emerges early in this portion of the observations and then remains strong for the entire segment.  We also note that the less significant peak of Segment 1 of the same sector may have reappeared in this segment, but again with low significance.
 
 \subsection{Sector 47}
\begin{figure}[h]
    \centering
    \resizebox{19cm}{!}{\includegraphics{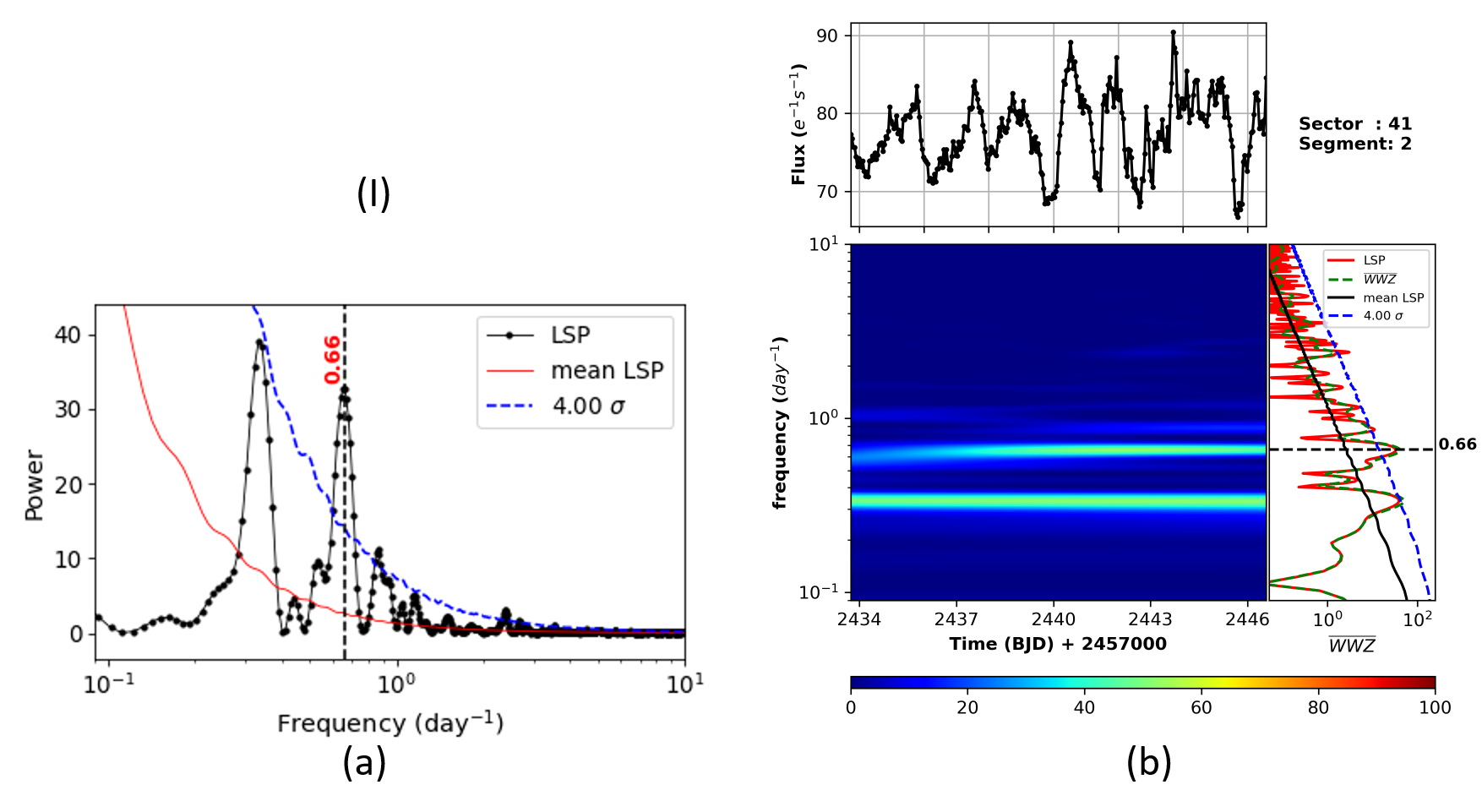}}
    \vspace{1cm}
    \resizebox{19cm}{!}{\includegraphics{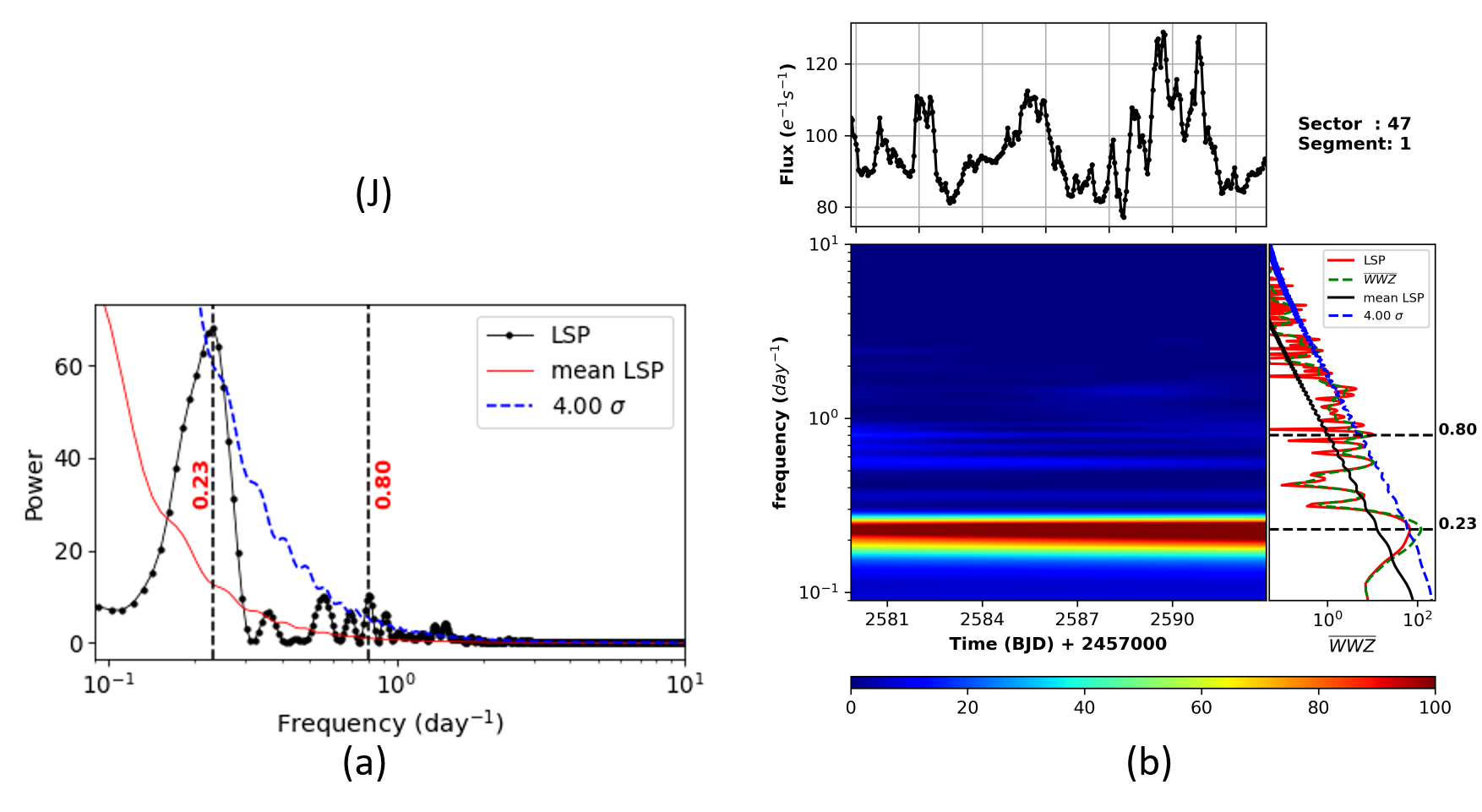}}
    \label{fig:Picture9}
\end{figure}

\begin{figure}[h]
    \centering
    \resizebox{19cm}{!}{\includegraphics{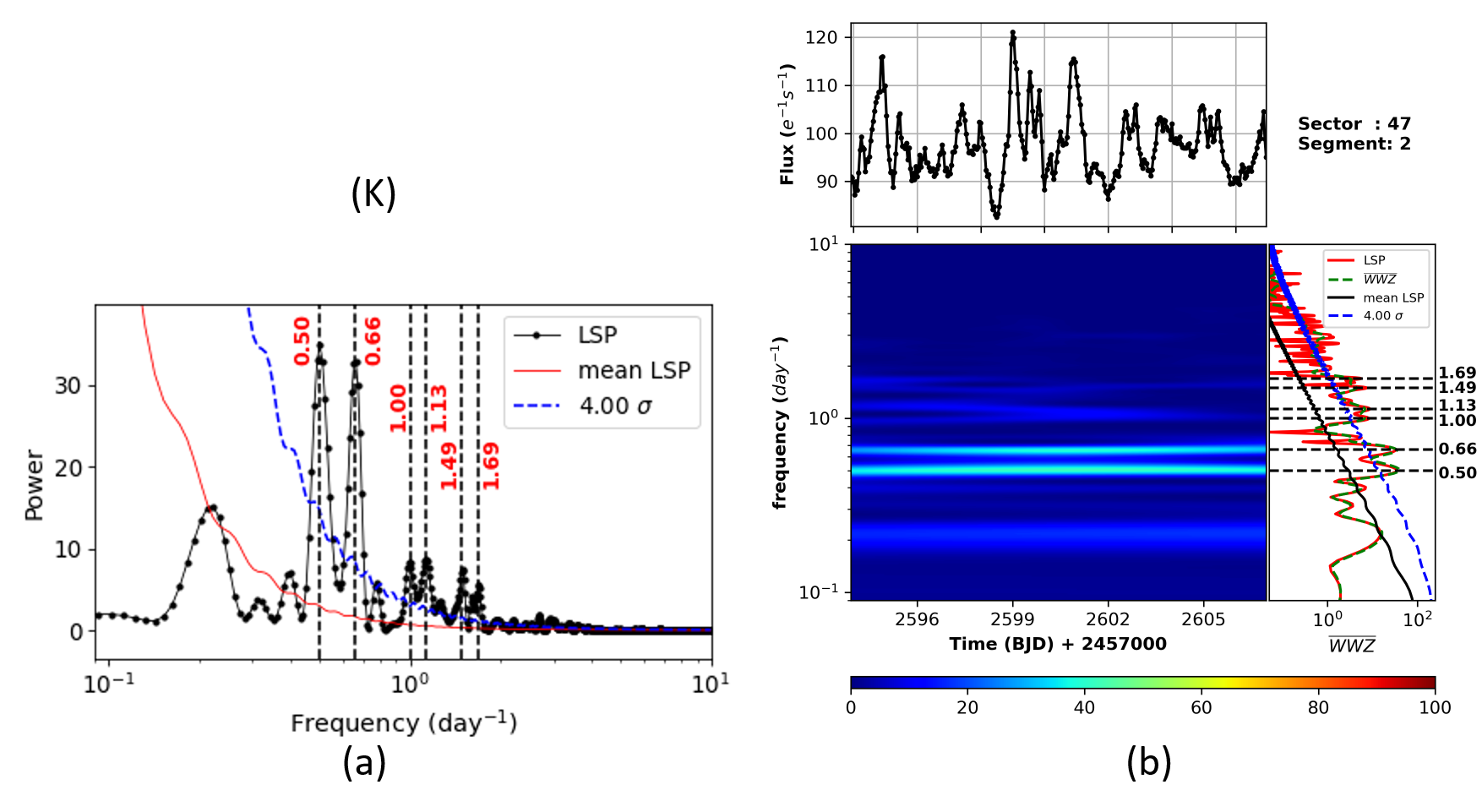}}
    \label{fig:Picture10}
\end{figure}

 \paragraph{Segment 1}
 Variations in this LC are strong, but provide no obvious signs of QPOs.  Nonetheless,
  two peaks are seen in Fig.\ \ref{fig:Picture9}J to rise above the 4.0$\sigma$ significance level, at frequencies of 0.23 day$^{-1}$ with appreciable GLSP and TAP powers and of 0.80 day$^{-1}$ with not  very high powers. Again, the longer period corresponding to  0.23 day$^{-1}$ does not meet our criteria for its validation. The period for  0.80 day$^{-1}$ or 1.25 days is thus the most likely QPO during this epoch, and it is present throughout the whole segment.
  \paragraph{Segment 2}
  The LC  analysis of this segment is particularly interesting.  As seen in Fig.\ \ref{fig:Picture10}K,  there are  several peaks with substantial and significant powers found in this segment. Two peaks at frequencies 0.50 day$^{-1}$ and 0.66 day$^{-1}$ have the greatest powers and cross the 10$\sigma$ significance level for both the GLSP and TAP analyses.  In addition to these frequency peaks, there are other peaks as well at frequencies: 1.00 day$^{-1}$, 1.13 day$^{-1}$, 1.49 day$^{-1}$ and 1.69 day$^{-1}$. These frequency peaks happen to be  rough multiples of the former two frequency peaks,  indicating the likely presence of higher-order harmonics of two  simultaneously significant peaks.  Though the periodicities at 0.50 day$^{-1}$ and 0.66 day$^{-1}$ persist throughout the segment, the rest of the peaks appear and disappear during the segment. We note that the 0.66 day$^{-1}$ QPO has the same frequency as seen in several other segments.\\
  
\subsection{Black hole mass estimation}  
  Along with its spin and charge, one of the most important quantities of any black hole (BH) is its mass and there are  several methods used  to estimate the central BH mass \citep[e.g.,][]{2000ApJ...540L..13P, 2002MNRAS.331..795M, 2003MNRAS.340..632L, 2005MNRAS.361..919P, 2013ApJ...773...90G}. These methods  for the BH mass estimation all require either measurement of spectral lines and/or the bolometric luminosity. Because  spectral lines  are absent in this source, and we lack its bolometric luminosity during our analysis, we have followed the method described in \citet{2009ApJ...690..216G} to  put limits on the SMBH mass for our source  under the assumption that the accrection disk around the SMBH is ultimately responsible for the observed QPO periods. The  limited range ($\sim$0.6--3 days) of periodic time-scales obtained in all the segments  imply that these variations  probably originate in the same general region. The minimum QPO time-scale found from our results can be used to estimate the mass of the central BH  within the range spanned by a spinless BH (Schwarzschild metric, $a = 0$)  and a very rapidly spinning BH (Kerr metric, ($a = 0.9982$).\\
\\If the period is assumed to be related to the orbital time-scale of some blob or flare near the inner region of the accretion disk \citep[e.g.][]{1991A&A...246...21Z} then using Eq.\ (4) of \citet{2009ApJ...690..216G} the  mass, $M$, of the SMBH is given by 
\begin{equation}
    \frac{M}{M_\odot}=\frac{3.23 \times 10^4\ P}{(r^{3/2}+a)(1+z)},
    \label{mass_period_rel}
\end{equation}

\noindent
 where $M_\odot$ is the sun's mass, $P$ is the observed period in seconds, $r$ is the radial distance, set to the last innermost stable circular orbit (ISCO),  in units of GM/c$^2$. 
 Using the minimum time-scale of 0.59 days (peak frequency  at 1.69 day$^{-1}$ from Sector 47 Segment 2) this leads to an SMBH mass of $(8.22\pm0.16)\times10^7M_\odot$ for a Schwarzschild
BH and $(5.22\pm0.10)\times10^8M_\odot$ for a Kerr BH), but it is possible that some emissions arise at distances within the last stable orbit  \citep[e.g.][]{1982Natur.300..506A}. Considering how often the period of 1.52 d is present, it is  more likely to  pertain to the ISCO.  Taking this into account, we suggest that within this framework, more probable BH masses   are $(2.12\pm0.06)\times10^8M_\odot$ for a Schwarzschild BH and  $(1.35\pm0.04\times10^9M_\odot)$  for a Kerr BH.\\

\section{Discussion \& Conclusion}
\noindent
In this work, we present the most extensive  search for optical QPOs  in multiple high cadence and evenly sampled datasets collected from the TESS satellite for the blazar S4 0954+658. The data presented here  include all measurements made from the launch of TESS  through Jan 2022. We have done a segment-wise analysis (described in Sect.\ \ref{Segmentation_DA}) of   the 6 different sector LCs of S4 0954+658. The results   are summarized in Table \ref{tab:Reult}. We have found multiple QPOs in various sectors with differing periodicities. The  period of the QPOs that we found mainly lie within  the range of 1 -- 3  days. Hints of both longer periods and occasional indications of shorter period harmonics are also found.  We have used both the robust generalized Lomb Scargle periodogram and the complementary method of wavelet analysis for the confirmation of any detected frequency peak. However we did not confirm and cannot further comment on the possible short-time-scale QPO at 0.62 days found by  \citet{2021MNRAS.504.5629R} that includes analyses of the first three sectors, however we do find a relatively weak signal of 0.59 days quasi-periodicity that has appreciable significance  in the second segment of Sector 47. Our segment-wise analysis of TESS data alone does not allow us to search for the longer period around 31 d suggested by \citet{2021MNRAS.504.5629R} based on composite light curves.\\
\\ The fact that the   frequency $\sim$0.66 day$^{-1}$  has been found to recur in various segments  of the different sectors, means that it  might be considered as  the dominant QPO  frequency for S4 0954+658. There are peaks at   frequencies 1.13 day$^{-1}$ and 1.69 day$^{-1}$ in Segment 2 of Sector 47, which are more precisely the harmonics of 0.56 day$^{-1}$ and not of 0.66 day$^{-1}$.   This suggests that  since 0.56 day$^{-1}$ is quite    close to 0.66 day$^{-1}$ in frequency space, and both frequency peaks were present at relatively close epochs, we may have observed  a  transition to the later frequency. The presence of both 0.56 day$^{-1}$ and 0.66 day$^{-1}$ peaks in the WWZ plots of  sector 41 also  support such a transition between the two frequencies.  The mechanism producing QPOs on time-scales of several hours could be of disk origin or could arise from relativistic jets. Possible disk explanations include location change of the dominant disk hotspot, small epicyclic deviations in both vertical and radial directions from exact planar motions within a thin accretion disk, and trapped pulsational modes within a disk \citep[e.g.,][and references therein]{1997ApJ...476..589P,2005AN....326..782A,2008ApJ...679..182E}. It is also possible for them to arise within the jet through a change in its diameter or growth of a kink instability \citep[e.g.,][]{2022Natur.609..265J}.\\
\\Several models based on binary SMBH systems \citep[e.g][]{2008Natur.452..851V,2010MNRAS.402.2087V,2015ApJ...813L..41A}, persistent jet precession and Lense-Thirring precession of accretion discs \citep[e.g.][]{1998ApJ...492L..59S,2000A&A...360...57R,2004ApJ...615L...5R,2018MNRAS.474L..81L}  have been proposed to explain QPOs in blazars on years-like periods. Since here, we detected QPOs with periods of a few days,  we can probably eliminate such models to explain the optical QPOs reported here in the blazar S4 0954+658.\\ 
\\The redshift ($z$) of the blazar S4 0954+658 as mentioned earlier is 0.3694, so the intrinsic  time-scales corresponding to the observed ones are reduced by the factor of $1+z$ (i.e., 1.3694), hence  shifting the source-frame period of variability in the range of 0.43 days to 3.65 days before considering Doppler factors.  Although such fast quasi-periodic variations can be produced by dominant turbulent cells, this mechanism is unlikely because of the multiple appearances of similar periodic time-scales \citep{1993A&A...271..344W}.
Shock-in-jet models  are known to dominate large-scale flares but are unlikely to produce essentially periodic fluctuations on observed time-scales much less than a year  \citep{1995ARA&A..33..163W}. \\
 \\The Doppler factor reported for the source lies in the range 6.1 $< \delta < $  35 \citep{2016PASJ...68...51T, 2017ApJ...846...98J, 2018A&A...617A..30M}. So, models involving jet twists or precession and other ways to change the viewing angle  combined with the Doppler boosting factor are also unlikely to provide valid explanations.   As noted earlier, their expected source frame time-scales are in years, so even with the maximum reported $\delta$, the observed time-scales might come down to  months or so, which are much longer than seen here for S4 0954+658.\\
 \\However, if these periodic perturbations in the inner part of the disk are advected into the jet so the observed emission comes from a relativistic flow directly affected by those perturbations, then the value of M would require the results obtained from Eq.\ (\ref{mass_period_rel}) to be multiplied by $\delta$, \citep{2009ApJ...690..216G}.   Unfortunately the range of published Doppler values for this source is broad (6.1 -- 35), but adopting the 1.52 d period as the fundamental one, we obtain the following ranges for the SMBH mass:  (1.3--$7.4) \times 10^9M_\odot$ for a Schwarzschild BH and (8.5--$47) \times 10^9M_\odot$ for  an extreme Kerr one.  The extremely high values arising from the combination of high Doppler factors and Kerr metric make them disfavored together. \\
\\
This paper includes data collected by the TESS mission. Funding for the TESS mission is provided by the NASA Explorer Program. S. Kishore thankfully acknowledges Mr. Abhradeep Roy for his valuable help while learning QPO data analysis techniques. A.C.G. is partially supported by the Ministry of Science and Technology of China (grant No.
2018YFA0404601) and the National Science Foundation of China (grants No. 11621303, 11835009, and 11973033). \\
\\
{\it Facility:}  Transiting Exoplanet Survey Satellite (TESS); {\it Softwares:} lightkurve \citep{2018ascl.soft12013L}, DELCgen \citep{2013MNRAS.433..907E}

\bibliography{ref}

\begin{thebibliography}{}
\expandafter\ifx\csname natexlab\endcsname\relax\def\natexlab#1{#1}\fi
\providecommand{\url}[1]{\href{#1}{#1}}
\providecommand{\dodoi}[1]{doi:~\href{http://doi.org/#1}{\nolinkurl{#1}}}
\providecommand{\doeprint}[1]{\href{http://ascl.net/#1}{\nolinkurl{http://ascl.net/#1}}}
\providecommand{\doarXiv}[1]{\href{https://arxiv.org/abs/#1}{\nolinkurl{https://arxiv.org/abs/#1}}}

\bibitem[{{Abdo} {et~al.}(2010){Abdo}, {Ackermann}, {Agudo}, {Ajello}, {Aller},
  {Aller}, {Angelakis}, {Arkharov}, {Axelsson}, {Bach}, {Baldini}, {Ballet},
  {Barbiellini}, {Bastieri}, {Baughman}, {Bechtol}, {Bellazzini}, {Benitez},
  {Berdyugin}, {Berenji}, {Blandford}, {Bloom}, {Boettcher}, {Bonamente},
  {Borgland}, {Bregeon}, {Brez}, {Brigida}, {Bruel}, {Burnett}, {Burrows},
  {Buson}, {Caliandro}, {Calzoletti}, {Cameron}, {Capalbi}, {Caraveo},
  {Carosati}, {Casandjian}, {Cavazzuti}, {Cecchi}, {{\c{C}}elik}, {Charles},
  {Chaty}, {Chekhtman}, {Chen}, {Chiang}, {Chincarini}, {Ciprini}, {Claus},
  {Cohen-Tanugi}, {Colafrancesco}, {Cominsky}, {Conrad}, {Costamante},
  {Cutini}, {D'ammando}, {Deitrick}, {D'Elia}, {Dermer}, {de Angelis}, {de
  Palma}, {Digel}, {Donnarumma}, {Silva}, {Drell}, {Dubois}, {Dultzin},
  {Dumora}, {Falcone}, {Farnier}, {Favuzzi}, {Fegan}, {Focke}, {Forn{\'e}},
  {Fortin}, {Frailis}, {Fuhrmann}, {Fukazawa}, {Funk}, {Fusco}, {G{\'o}mez},
  {Gargano}, {Gasparrini}, {Gehrels}, {Germani}, {Giebels}, {Giglietto},
  {Giommi}, {Giordano}, {Giuliani}, {Glanzman}, {Godfrey}, {Grenier},
  {Gronwall}, {Grove}, {Guillemot}, {Guiriec}, {Gurwell}, {Hadasch},
  {Hanabata}, {Harding}, {Hayashida}, {Hays}, {Healey}, {Heidt}, {Hiriart},
  {Horan}, {Hoversten}, {Hughes}, {Itoh}, {Jackson}, {J{\'o}hannesson},
  {Johnson}, {Johnson}, {Jorstad}, {Kadler}, {Kamae}, {Katagiri}, {Kataoka},
  {Kawai}, {Kennea}, {Kerr}, {Kimeridze}, {Kn{\"o}dlseder}, {Kocian},
  {Kopatskaya}, {Koptelova}, {Konstantinova}, {Kovalev}, {Kovalev},
  {Kurtanidze}, {Kuss}, {Lande}, {Larionov}, {Latronico}, {Leto}, {Lindfors},
  {Longo}, {Loparco}, {Lott}, {Lovellette}, {Lubrano}, {Madejski}, {Makeev},
  {Marchegiani}, {Marscher}, {Marshall}, {Max-Moerbeck}, {Mazziotta},
  {McConville}, {McEnery}, {Meurer}, {Michelson}, {Mitthumsiri}, {Mizuno},
  {Moiseev}, {Monte}, {Monzani}, {Morselli}, {Moskalenko}, {Murgia},
  {Nestoras}, {Nilsson}, {Nizhelsky}, {Nolan}, {Norris}, {Nuss}, {Ohsugi},
  {Ojha}, {Omodei}, {Orlando}, {Ormes}, {Osborne}, {Ozaki}, {Pacciani},
  {Padovani}, {Pagani}, {Page}, {Paneque}, {Panetta}, {Parent}, {Pasanen},
  {Pavlidou}, {Pelassa}, {Pepe}, {Perri}, {Pesce-Rollins}, {Piranomonte},
  {Piron}, {Pittori}, {Porter}, {Puccetti}, {Rahoui}, {Rain{\`o}}, {Raiteri},
  {Rando}, {Razzano}, {Reimer}, {Reimer}, {Reposeur}, {Richards}, {Ritz},
  {Rochester}, {Rodriguez}, {Romani}, {Ros}, {Roth}, {Roustazadeh}, {Ryde},
  {Sadrozinski}, {Sadun}, {Sanchez}, {Sander}, {Saz Parkinson}, {Scargle},
  {Sellerholm}, {Sgr{\`o}}, {Shaw}, {Sigua}, {Siskind}, {Smith}, {Smith},
  {Spandre}, {Spinelli}, {Starck}, {Stevenson}, {Stratta}, {Strickman},
  {Suson}, {Tajima}, {Takahashi}, {Takahashi}, {Takalo}, {Tanaka}, {Thayer},
  {Thayer}, {Thompson}, {Tibaldo}, {Torres}, {Tosti}, {Tramacere}, {Uchiyama},
  {Usher}, {Vasileiou}, {Verrecchia}, {Vilchez}, {Villata}, {Vitale}, {Waite},
  {Wang}, {Winer}, {Wood}, {Ylinen}, {Zensus}, {Zhekanis}, \&
  {Ziegler}}]{2010ApJ...716...30A}
{Abdo}, A.~A., {Ackermann}, M., {Agudo}, I., {et~al.} 2010, \apj, 716, 30,
  \dodoi{10.1088/0004-637X/716/1/30}

\bibitem[{{Abramowicz}(2005)}]{2005AN....326..782A}
{Abramowicz}, M.~A. 2005, Astronomische Nachrichten, 326, 782,
  \dodoi{10.1002/asna.200510413}

\bibitem[{{Abramowicz} \& {Nobili}(1982)}]{1982Natur.300..506A}
{Abramowicz}, M.~A., \& {Nobili}, L. 1982, \nat, 300, 506,
  \dodoi{10.1038/300506a0}

\bibitem[{{Ackermann} {et~al.}(2015){Ackermann}, {Ajello}, {Albert}, {Atwood},
  {Baldini}, {Ballet}, {Barbiellini}, {Bastieri}, {Becerra Gonzalez},
  {Bellazzini}, {Bissaldi}, {Blandford}, {Bloom}, {Bonino}, {Bottacini},
  {Bregeon}, {Bruel}, {Buehler}, {Buson}, {Caliandro}, {Cameron}, {Caputo},
  {Caragiulo}, {Caraveo}, {Cavazzuti}, {Cecchi}, {Chekhtman}, {Chiang},
  {Chiaro}, {Ciprini}, {Cohen-Tanugi}, {Conrad}, {Cutini}, {D'Ammando}, {de
  Angelis}, {de Palma}, {Desiante}, {Di Venere}, {Dom{\'\i}nguez}, {Drell},
  {Favuzzi}, {Fegan}, {Ferrara}, {Focke}, {Fuhrmann}, {Fukazawa}, {Fusco},
  {Gargano}, {Gasparrini}, {Giglietto}, {Giommi}, {Giordano}, {Giroletti},
  {Godfrey}, {Green}, {Grenier}, {Grove}, {Guiriec}, {Harding}, {Hays},
  {Hewitt}, {Hill}, {Horan}, {Jogler}, {J{\'o}hannesson}, {Johnson}, {Kamae},
  {Kuss}, {Larsson}, {Latronico}, {Li}, {Li}, {Longo}, {Loparco}, {Lott},
  {Lovellette}, {Lubrano}, {Magill}, {Maldera}, {Manfreda}, {Max-Moerbeck},
  {Mayer}, {Mazziotta}, {McEnery}, {Michelson}, {Mizuno}, {Monzani},
  {Morselli}, {Moskalenko}, {Murgia}, {Nuss}, {Ohno}, {Ohsugi}, {Ojha},
  {Omodei}, {Orlando}, {Ormes}, {Paneque}, {Pearson}, {Perkins}, {Perri},
  {Pesce-Rollins}, {Petrosian}, {Piron}, {Pivato}, {Porter}, {Rain{\`o}},
  {Rando}, {Razzano}, {Readhead}, {Reimer}, {Reimer}, {Schulz}, {Sgr{\`o}},
  {Siskind}, {Spada}, {Spandre}, {Spinelli}, {Suson}, {Takahashi}, {Thayer},
  {Thompson}, {Tibaldo}, {Torres}, {Tosti}, {Troja}, {Uchiyama}, {Vianello},
  {Wood}, {Wood}, {Zimmer}, {Berdyugin}, {Corbet}, {Hovatta}, {Lindfors},
  {Nilsson}, {Reinthal}, {Sillanp{\"a}{\"a}}, {Stamerra}, {Takalo}, \&
  {Valtonen}}]{2015ApJ...813L..41A}
{Ackermann}, M., {Ajello}, M., {Albert}, A., {et~al.} 2015, \apjl, 813, L41,
  \dodoi{10.1088/2041-8205/813/2/L41}

\bibitem[{{Alston} {et~al.}(2014){Alston}, {Markeviciute}, {Kara}, {Fabian}, \&
  {Middleton}}]{2014MNRAS.445L..16A}
{Alston}, W.~N., {Markeviciute}, J., {Kara}, E., {Fabian}, A.~C., \&
  {Middleton}, M. 2014, \mnras, 445, L16, \dodoi{10.1093/mnrasl/slu127}

\bibitem[{{Alston} {et~al.}(2015){Alston}, {Parker},
  {Markevi{\v{c}}i{\={u}}t{\.{e}}}, {Fabian}, {Middleton}, {Lohfink}, {Kara},
  \& {Pinto}}]{2015MNRAS.449..467A}
{Alston}, W.~N., {Parker}, M.~L., {Markevi{\v{c}}i{\={u}}t{\.{e}}}, J.,
  {et~al.} 2015, \mnras, 449, 467, \dodoi{10.1093/mnras/stv351}

\bibitem[{{Bachev}(2015)}]{2015MNRAS.451L..21B}
{Bachev}, R. 2015, \mnras, 451, L21, \dodoi{10.1093/mnrasl/slv059}

\bibitem[{{Becerra Gonz{\'a}lez} {et~al.}(2021){Becerra Gonz{\'a}lez},
  {Acosta-Pulido}, {Boschin}, {Clavero}, {Otero-Santos}, {Carballo-Bello}, \&
  {Dom{\'\i}nguez-Palmero}}]{2021MNRAS.504.5258B}
{Becerra Gonz{\'a}lez}, J., {Acosta-Pulido}, J.~A., {Boschin}, W., {et~al.}
  2021, \mnras, 504, 5258, \dodoi{10.1093/mnras/stab1274}

\bibitem[{{Bhatta}(2017)}]{2017ApJ...847....7B}
{Bhatta}, G. 2017, \apj, 847, 7, \dodoi{10.3847/1538-4357/aa86ed}

\bibitem[{Bhatta(2019)}]{2019MNRAS.487.3990B}
Bhatta, G. 2019, \mnras, 487, 3990, \dodoi{10.1093/mnras/stz1482}

\bibitem[{{Blandford} \& {Rees}(1978)}]{1978PhyS...17..265B}
{Blandford}, R.~D., \& {Rees}, M.~J. 1978, \physscr, 17, 265,
  \dodoi{10.1088/0031-8949/17/3/020}

\bibitem[{{Carini} {et~al.}(2020){Carini}, {Wehrle}, {Wiita}, {Ward}, \&
  {Pendleton}}]{2020ApJ...903..134C}
{Carini}, M., {Wehrle}, A.~E., {Wiita}, P.~J., {Ward}, Z., \& {Pendleton}, K.
  2020, \apj, 903, 134, \dodoi{10.3847/1538-4357/abbb92}

\bibitem[{{Emmanoulopoulos} {et~al.}(2013){Emmanoulopoulos}, {McHardy}, \&
  {Papadakis}}]{2013MNRAS.433..907E}
{Emmanoulopoulos}, D., {McHardy}, I.~M., \& {Papadakis}, I.~E. 2013, \mnras,
  433, 907, \dodoi{10.1093/mnras/stt764}

\bibitem[{{Espaillat} {et~al.}(2008){Espaillat}, {Bregman}, {Hughes}, \&
  {Lloyd-Davies}}]{2008ApJ...679..182E}
{Espaillat}, C., {Bregman}, J., {Hughes}, P., \& {Lloyd-Davies}, E. 2008, \apj,
  679, 182, \dodoi{10.1086/587023}

\bibitem[{{Foster}(1996)}]{1996AJ....112.1709F}
{Foster}, G. 1996, \aj, 112, 1709, \dodoi{10.1086/118137}

\bibitem[{{Gaur} {et~al.}(2010){Gaur}, {Gupta}, {Lachowicz}, \&
  {Wiita}}]{2010ApJ...718..279G}
{Gaur}, H., {Gupta}, A.~C., {Lachowicz}, P., \& {Wiita}, P.~J. 2010, \apj, 718,
  279, \dodoi{10.1088/0004-637X/718/1/279}

\bibitem[{{Ghisellini} {et~al.}(2011){Ghisellini}, {Tavecchio}, {Foschini}, \&
  {Ghirlanda}}]{2011MNRAS.414.2674G}
{Ghisellini}, G., {Tavecchio}, F., {Foschini}, L., \& {Ghirlanda}, G. 2011,
  \mnras, 414, 2674, \dodoi{10.1111/j.1365-2966.2011.18578.x}

\bibitem[{{Ghisellini} {et~al.}(1997){Ghisellini}, {Villata}, {Raiteri},
  {Bosio}, {de Francesco}, {Latini}, {Maesano}, {Massaro}, {Montagni}, {Nesci},
  {Tosti}, {Fiorucci}, {Pian}, {Maraschi}, {Treves}, {Comastri}, \&
  {Mignoli}}]{1997A&A...327...61G}
{Ghisellini}, G., {Villata}, M., {Raiteri}, C.~M., {et~al.} 1997, \aap, 327,
  61.
\newblock \doarXiv{astro-ph/9706254}

\bibitem[{{Gierli{\'n}ski} {et~al.}(2008){Gierli{\'n}ski}, {Middleton}, {Ward},
  \& {Done}}]{2008Natur.455..369G}
{Gierli{\'n}ski}, M., {Middleton}, M., {Ward}, M., \& {Done}, C. 2008, \nat,
  455, 369, \dodoi{10.1038/nature07277}

\bibitem[{{Grier} {et~al.}(2013){Grier}, {Martini}, {Watson}, {Peterson},
  {Bentz}, {Dasyra}, {Dietrich}, {Ferrarese}, {Pogge}, \&
  {Zu}}]{2013ApJ...773...90G}
{Grier}, C.~J., {Martini}, P., {Watson}, L.~C., {et~al.} 2013, \apj, 773, 90,
  \dodoi{10.1088/0004-637X/773/2/90}

\bibitem[{{Gupta}(2018)}]{2018Galax...6....1G}
{Gupta}, A. 2018, Galaxies, 6, 1, \dodoi{10.3390/galaxies6010001}

\bibitem[{{Gupta}(2014)}]{2014JApA...35..307G}
{Gupta}, A.~C. 2014, Journal of Astrophysics and Astronomy, 35, 307,
  \dodoi{10.1007/s12036-014-9219-7}

\bibitem[{{Gupta} {et~al.}(2009){Gupta}, {Srivastava}, \&
  {Wiita}}]{2009ApJ...690..216G}
{Gupta}, A.~C., {Srivastava}, A.~K., \& {Wiita}, P.~J. 2009, \apj, 690, 216,
  \dodoi{10.1088/0004-637X/690/1/216}

\bibitem[{{Gupta} {et~al.}(2018){Gupta}, {Tripathi}, {Wiita}, {Gu}, {Bambi}, \&
  {Ho}}]{2018A&A...616L...6G}
{Gupta}, A.~C., {Tripathi}, A., {Wiita}, P.~J., {et~al.} 2018, \aap, 616, L6,
  \dodoi{10.1051/0004-6361/201833629}

\bibitem[{{Gupta} {et~al.}(2019{\natexlab{a}}){Gupta}, {Tripathi}, {Wiita},
  {Kushwaha}, {Zhang}, \& {Bambi}}]{2019MNRAS.484.5785G}
---. 2019{\natexlab{a}}, \mnras, 484, 5785, \dodoi{10.1093/mnras/stz395}

\bibitem[{{Gupta} {et~al.}(2019{\natexlab{b}}){Gupta}, {Gaur}, {Wiita},
  {Pandey}, {Kushwaha}, {Hu}, {Kurtanidze}, {Semkov}, {Damljanovic}, {Goyal},
  {Uemura}, {Darriba}, {Chen}, {Vince}, {Gu}, {Zhang}, {Bachev}, {Chanishvili},
  {Itoh}, {Kawabata}, {Kurtanidze}, {Nakaoka}, {Nikolashvili}, {Stawarz}, \&
  {Strigachev}}]{2019AJ....157...95G}
{Gupta}, A.~C., {Gaur}, H., {Wiita}, P.~J., {et~al.} 2019{\natexlab{b}}, \aj,
  157, 95, \dodoi{10.3847/1538-3881/aafe7d}

\bibitem[{{Hagen-Thorn} {et~al.}(2015){Hagen-Thorn}, {Larionov}, {Arkharov},
  {Hagen-Thorn}, {Blinov}, {Morozova}, {Troitskaya}, {Takalo}, \&
  {Sillanpy{\"a}{\"a}}}]{2015ARep...59..551H}
{Hagen-Thorn}, V.~A., {Larionov}, V.~M., {Arkharov}, A.~A., {et~al.} 2015,
  Astronomy Reports, 59, 551, \dodoi{10.1134/S1063772915050030}

\bibitem[{{Hervet} {et~al.}(2016){Hervet}, {Boisson}, \&
  {Sol}}]{2016A&A...592A..22H}
{Hervet}, O., {Boisson}, C., \& {Sol}, H. 2016, \aap, 592, A22,
  \dodoi{10.1051/0004-6361/201628117}

\bibitem[{{Horne} \& {Baliunas}(1986)}]{1986ApJ...302..757H}
{Horne}, J.~H., \& {Baliunas}, S.~L. 1986, \apj, 302, 757,
  \dodoi{10.1086/164037}

\bibitem[{James {et~al.}(2013)James, Witten, Hastie, \&
  Tibshirani}]{james2013introduction}
James, G., Witten, D., Hastie, T., \& Tibshirani, R. 2013, An introduction to
  statistical learning (Springer), \dodoi{10.1007/978-1-4614-7138-7}

\bibitem[{{Jenkins} {et~al.}(2016){Jenkins}, {Twicken}, {McCauliff},
  {Campbell}, {Sanderfer}, {Lung}, {Mansouri-Samani}, {Girouard}, {Tenenbaum},
  {Klaus}, {Smith}, {Caldwell}, {Chacon}, {Henze}, {Heiges}, {Latham},
  {Morgan}, {Swade}, {Rinehart}, \& {Vanderspek}}]{2016SPIE.9913E..3EJ}
{Jenkins}, J.~M., {Twicken}, J.~D., {McCauliff}, S., {et~al.} 2016, in Society
  of Photo-Optical Instrumentation Engineers (SPIE) Conference Series, Vol.
  9913, Software and Cyberinfrastructure for Astronomy IV, ed. G.~{Chiozzi} \&
  J.~C. {Guzman}, 99133E, \dodoi{10.1117/12.2233418}

\bibitem[{{Jorstad} {et~al.}(2017){Jorstad}, {Marscher}, {Morozova},
  {Troitsky}, {Agudo}, {Casadio}, {Foord}, {G{\'o}mez}, {MacDonald}, {Molina},
  {L{\"a}hteenm{\"a}ki}, {Tammi}, \& {Tornikoski}}]{2017ApJ...846...98J}
{Jorstad}, S.~G., {Marscher}, A.~P., {Morozova}, D.~A., {et~al.} 2017, \apj,
  846, 98, \dodoi{10.3847/1538-4357/aa8407}

\bibitem[{{Jorstad} {et~al.}(2022){Jorstad}, {Marscher}, {Raiteri}, {Villata},
  {Weaver}, {Zhang}, {Dong}, {G{\'o}mez}, {Perel}, {Savchenko}, {Larionov},
  {Carosati}, {Chen}, {Kurtanidze}, {Marchini}, {Matsumoto}, {Mortari},
  {Aceti}, {Acosta-Pulido}, {Andreeva}, {Apolonio}, {Arena}, {Arkharov},
  {Bachev}, {Banfi}, {Bonnoli}, {Borman}, {Bozhilov}, {Carnerero},
  {Damljanovic}, {Ehgamberdiev}, {Els{\"a}sser}, {Frasca}, {Gabellini},
  {Grishina}, {Gupta}, {Hagen-Thorn}, {Hallum}, {Hart}, {Hasuda}, {Hemrich},
  {Hsiao}, {Ibryamov}, {Irsmambetova}, {Ivanov}, {Joner}, {Kimeridze},
  {Klimanov}, {Kn{\"o}tt}, {Kopatskaya}, {Kurtanidze}, {Kurtenkov}, {Kuutma},
  {Larionova}, {Leonini}, {Lin}, {Lorey}, {Mannheim}, {Marino}, {Minev},
  {Mirzaqulov}, {Morozova}, {Nikiforova}, {Nikolashvili}, {Ovcharov}, {Papini},
  {Pursimo}, {Rahimov}, {Reinhart}, {Sakamoto}, {Salvaggio}, {Semkov},
  {Shakhovskoy}, {Sigua}, {Steineke}, {Stojanovic}, {Strigachev}, {Troitskaya},
  {Troitskiy}, {Tsai}, {Valcheva}, {Vasilyev}, {Vince}, {Waller}, {Zaharieva},
  \& {Chatterjee}}]{2022Natur.609..265J}
{Jorstad}, S.~G., {Marscher}, A.~P., {Raiteri}, C.~M., {et~al.} 2022, \nat,
  609, 265, \dodoi{10.1038/s41586-022-05038-9}

\bibitem[{{Kollgaard} {et~al.}(1992){Kollgaard}, {Wardle}, {Roberts}, \&
  {Gabuzda}}]{1992AJ....104.1687K}
{Kollgaard}, R.~I., {Wardle}, J.~F.~C., {Roberts}, D.~H., \& {Gabuzda}, D.~C.
  1992, \aj, 104, 1687, \dodoi{10.1086/116352}

\bibitem[{{Kushwaha} {et~al.}(2020){Kushwaha}, {Sarkar}, {Gupta}, {Tripathi},
  \& {Wiita}}]{2020MNRAS.499..653K}
{Kushwaha}, P., {Sarkar}, A., {Gupta}, A.~C., {Tripathi}, A., \& {Wiita}, P.~J.
  2020, \mnras, 499, 653, \dodoi{10.1093/mnras/staa2899}

\bibitem[{{Lachowicz} {et~al.}(2009){Lachowicz}, {Gupta}, {Gaur}, \&
  {Wiita}}]{2009A&A...506L..17L}
{Lachowicz}, P., {Gupta}, A.~C., {Gaur}, H., \& {Wiita}, P.~J. 2009, \aap, 506,
  L17, \dodoi{10.1051/0004-6361/200913161}

\bibitem[{{Liang} \& {Liu}(2003)}]{2003MNRAS.340..632L}
{Liang}, E.~W., \& {Liu}, H.~T. 2003, \mnras, 340, 632,
  \dodoi{10.1046/j.1365-8711.2003.06327.x}

\bibitem[{{Lightkurve Collaboration} {et~al.}(2018){Lightkurve Collaboration},
  {Cardoso}, {Hedges}, {Gully-Santiago}, {Saunders}, {Cody}, {Barclay}, {Hall},
  {Sagear}, {Turtelboom}, {Zhang}, {Tzanidakis}, {Mighell}, {Coughlin}, {Bell},
  {Berta-Thompson}, {Williams}, {Dotson}, \& {Barentsen}}]{2018ascl.soft12013L}
{Lightkurve Collaboration}, {Cardoso}, J.~V.~d.~M., {Hedges}, C., {et~al.}
  2018, {Lightkurve: Kepler and TESS time series analysis in Python},
  Astrophysics Source Code Library.
\newblock \doeprint{1812.013}

\bibitem[{{Lin} {et~al.}(2013){Lin}, {Irwin}, {Godet}, {Webb}, \&
  {Barret}}]{2013ApJ...776L..10L}
{Lin}, D., {Irwin}, J.~A., {Godet}, O., {Webb}, N.~A., \& {Barret}, D. 2013,
  \apjl, 776, L10, \dodoi{10.1088/2041-8205/776/1/L10}

\bibitem[{{Liska} {et~al.}(2018){Liska}, {Hesp}, {Tchekhovskoy}, {Ingram}, {van
  der Klis}, \& {Markoff}}]{2018MNRAS.474L..81L}
{Liska}, M., {Hesp}, C., {Tchekhovskoy}, A., {et~al.} 2018, \mnras, 474, L81,
  \dodoi{10.1093/mnrasl/slx174}

\bibitem[{{Lomb}(1976)}]{1976Ap&SS..39..447L}
{Lomb}, N.~R. 1976, \apss, 39, 447, \dodoi{10.1007/BF00648343}

\bibitem[{{MAGIC Collaboration} {et~al.}(2018){MAGIC Collaboration}, {Ahnen},
  {Ansoldi}, {Antonelli}, {Arcaro}, {Baack}, {Babi{\'c}}, {Banerjee},
  {Bangale}, {Barres de Almeida}, {Barrio}, {Bednarek}, {Bernardini}, {Berse},
  {Berti}, {Bhattacharyya}, {Biland}, {Blanch}, {Bonnoli}, {Carosi}, {Carosi},
  {Ceribella}, {Chatterjee}, {Colak}, {Colin}, {Colombo}, {Contreras},
  {Cortina}, {Covino}, {Cumani}, {da Vela}, {Dazzi}, {de Angelis}, {de Lotto},
  {Delfino}, {Delgado}, {di Pierro}, {Dom{\'\i}nguez}, {Dominis Prester},
  {Dorner}, {Doro}, {Einecke}, {Elsaesser}, {Fallah Ramazani},
  {Fern{\'a}ndez-Barral}, {Fidalgo}, {Fonseca}, {Font}, {Fruck}, {Galindo},
  {Garc{\'\i}a L{\'o}pez}, {Garczarczyk}, {Gaug}, {Giammaria}, {Godinovi{\'c}},
  {Gora}, {Guberman}, {Hadasch}, {Hahn}, {Hassan}, {Hayashida}, {Herrera},
  {Hose}, {Hrupec}, {Ishio}, {Konno}, {Kubo}, {Kushida}, {Kuve{\v{z}}di{\'c}},
  {Lelas}, {Lindfors}, {Lombardi}, {Longo}, {L{\'o}pez}, {Maggio}, {Majumdar},
  {Makariev}, {Maneva}, {Manganaro}, {Mannheim}, {Maraschi}, {Mariotti},
  {Mart{\'\i}nez}, {Masuda}, {Mazin}, {Mielke}, {Minev}, {Miranda}, {Mirzoyan},
  {Moralejo}, {Moreno}, {Moretti}, {Nagayoshi}, {Neustroev}, {Niedzwiecki},
  {Nievas Rosillo}, {Nigro}, {Nilsson}, {Ninci}, {Nishijima}, {Noda},
  {Nogu{\'e}s}, {Paiano}, {Palacio}, {Paneque}, {Paoletti}, {Paredes},
  {Pedaletti}, {Peresano}, {Persic}, {Prada Moroni}, {Prandini}, {Puljak},
  {Garcia}, {Reichardt}, {Rhode}, {Rib{\'o}}, {Rico}, {Righi}, {Rugliancich},
  {Saito}, {Satalecka}, {Schweizer}, {Sitarek}, {{\v{S}}nidari{\'c}},
  {Sobczynska}, {Stamerra}, {Strzys}, {Suri{\'c}}, {Takahashi}, {Takalo},
  {Tavecchio}, {Temnikov}, {Terzi{\'c}}, {Teshima}, {Torres-Alb{\`a}},
  {Treves}, {Tsujimoto}, {Vanzo}, {Vazquez Acosta}, {Vovk}, {Ward}, {Will},
  {Zari{\'c}}, {Becerra Gonz{\'a}lez}, {Tanaka}, {Ojha}, {Finke},
  {L{\"a}hteenm{\"a}ki}, {J{\"a}rvel{\"a}}, {Tornikoski}, {Ramakrishnan},
  {Hovatta}, {Jorstad}, {Marscher}, {Larionov}, {Borman}, {Grishina},
  {Kopatskaya}, {Larionova}, {Morozova}, {Savchenko}, {Troitskaya}, {Troitsky},
  {Vasilyev}, {Agudo}, {Molina}, {Casadio}, {Gurwell}, {Carnerero}, {Protasio},
  \& {Acosta Pulido}}]{2018A&A...617A..30M}
{MAGIC Collaboration}, {Ahnen}, M.~L., {Ansoldi}, S., {et~al.} 2018, \aap, 617,
  A30, \dodoi{10.1051/0004-6361/201832624}

\bibitem[{{Marcha} {et~al.}(1996){Marcha}, {Browne}, {Impey}, \&
  {Smith}}]{1996MNRAS.281..425M}
{Marcha}, M.~J.~M., {Browne}, I.~W.~A., {Impey}, C.~D., \& {Smith}, P.~S. 1996,
  \mnras, 281, 425, \dodoi{10.1093/mnras/281.2.425}

\bibitem[{{McLure} \& {Dunlop}(2002)}]{2002MNRAS.331..795M}
{McLure}, R.~J., \& {Dunlop}, J.~S. 2002, \mnras, 331, 795,
  \dodoi{10.1046/j.1365-8711.2002.05236.x}

\bibitem[{{Pan} {et~al.}(2016){Pan}, {Yuan}, {Yao}, {Zhou}, {Liu}, {Zhou}, \&
  {Zhang}}]{2016ApJ...819L..19P}
{Pan}, H.-W., {Yuan}, W., {Yao}, S., {et~al.} 2016, \apjl, 819, L19,
  \dodoi{10.3847/2041-8205/819/2/L19}

\bibitem[{{Perez} {et~al.}(1997){Perez}, {Silbergleit}, {Wagoner}, \&
  {Lehr}}]{1997ApJ...476..589P}
{Perez}, C.~A., {Silbergleit}, A.~S., {Wagoner}, R.~V., \& {Lehr}, D.~E. 1997,
  \apj, 476, 589, \dodoi{10.1086/303658}

\bibitem[{{Peterson} \& {Wandel}(2000)}]{2000ApJ...540L..13P}
{Peterson}, B.~M., \& {Wandel}, A. 2000, \apjl, 540, L13,
  \dodoi{10.1086/312862}

\bibitem[{{Pian} {et~al.}(2005){Pian}, {Falomo}, \&
  {Treves}}]{2005MNRAS.361..919P}
{Pian}, E., {Falomo}, R., \& {Treves}, A. 2005, \mnras, 361, 919,
  \dodoi{10.1111/j.1365-2966.2005.09216.x}

\bibitem[{{Press}(1978)}]{1978ComAp...7..103P}
{Press}, W.~H. 1978, Comments on Astrophysics, 7, 103

\bibitem[{Press {et~al.}(2007)Press, Teukolsky, Vetterling, \&
  Flannery}]{2007nrca.book.....P}
Press, W.~H., Teukolsky, S.~A., Vetterling, W.~T., \& Flannery, B.~P. 2007,
  Numerical Recipes 3rd Edition: The Art of Scientific Computing, 3rd edn.
  (USA: Cambridge University Press)

\bibitem[{{Raiteri} {et~al.}(1999){Raiteri}, {Villata}, {Tosti}, {Fiorucci},
  {Ghisellini}, {Takalo}, {Sillanp{\"a}{\"a}}, {Valtaoja}, {Ter{\"a}sranta},
  {Tornikoski}, {Aller}, {Aller}, {De Francesco}, {Hein{\"a}m{\"a}ki},
  {Katajainen}, {Lanteri}, {Nilsson}, {Pursimo}, {Rizzi}, \&
  {Sobrito}}]{1999A&A...352...19R}
{Raiteri}, C.~M., {Villata}, M., {Tosti}, G., {et~al.} 1999, \aap, 352, 19

\bibitem[{{Raiteri} {et~al.}(2021){Raiteri}, {Villata}, {Larionov}, {Jorstad},
  {Marscher}, {Weaver}, {Acosta-Pulido}, {Agudo}, {Andreeva}, {Arkharov},
  {Bachev}, {Ben{\'\i}tez}, {Berton}, {Bj{\"o}rklund}, {Borman}, {Bozhilov},
  {Carnerero}, {Carosati}, {Casadio}, {Chen}, {Damljanovic}, {D'Ammando},
  {Escudero}, {Fuentes}, {Giroletti}, {Grishina}, {Gupta}, {Hagen-Thorn},
  {Hart}, {Hiriart}, {Hou}, {Ivanov}, {Kim}, {Kimeridze}, {Konstantopoulou},
  {Kopatskaya}, {Kurtanidze}, {Kurtanidze}, {L{\"a}hteenm{\"a}ki}, {Larionova},
  {Larionova}, {Marchili}, {Markovic}, {Minev}, {Morozova}, {Myserlis},
  {Nakamura}, {Nikiforova}, {Nikolashvili}, {Otero-Santos}, {Ovcharov},
  {Pursimo}, {Rahimov}, {Righini}, {Sakamoto}, {Savchenko}, {Semkov},
  {Shakhovskoy}, {Sigua}, {Stojanovic}, {Strigachev}, {Thum}, {Tornikoski},
  {Traianou}, {Troitskaya}, {Troitskiy}, {Tsai}, {Valcheva}, {Vasilyev},
  {Vince}, \& {Zaharieva}}]{2021MNRAS.504.5629R}
{Raiteri}, C.~M., {Villata}, M., {Larionov}, V.~M., {et~al.} 2021, \mnras, 504,
  5629, \dodoi{10.1093/mnras/stab1268}

\bibitem[{{Rani} {et~al.}(2010){Rani}, {Gupta}, {Joshi}, {Ganesh}, \&
  {Wiita}}]{2010ApJ...719L.153R}
{Rani}, B., {Gupta}, A.~C., {Joshi}, U.~C., {Ganesh}, S., \& {Wiita}, P.~J.
  2010, \apjl, 719, L153, \dodoi{10.1088/2041-8205/719/2/L153}

\bibitem[{{Rani} {et~al.}(2009){Rani}, {Wiita}, \&
  {Gupta}}]{2009ApJ...696.2170R}
{Rani}, B., {Wiita}, P.~J., \& {Gupta}, A.~C. 2009, \apj, 696, 2170,
  \dodoi{10.1088/0004-637X/696/2/2170}

\bibitem[{{Remillard} \& {McClintock}(2006)}]{2006ARA&A..44...49R}
{Remillard}, R.~A., \& {McClintock}, J.~E. 2006, \araa, 44, 49,
  \dodoi{10.1146/annurev.astro.44.051905.092532}

\bibitem[{{Ricker} {et~al.}(2014){Ricker}, {Winn}, {Vanderspek}, {Latham},
  {Bakos}, {Bean}, {Berta-Thompson}, {Brown}, {Buchhave}, {Butler}, {Butler},
  {Chaplin}, {Charbonneau}, {Christensen-Dalsgaard}, {Clampin}, {Deming},
  {Doty}, {De Lee}, {Dressing}, {Dunham}, {Endl}, {Fressin}, {Ge}, {Henning},
  {Holman}, {Howard}, {Ida}, {Jenkins}, {Jernigan}, {Johnson}, {Kaltenegger},
  {Kawai}, {Kjeldsen}, {Laughlin}, {Levine}, {Lin}, {Lissauer}, {MacQueen},
  {Marcy}, {McCullough}, {Morton}, {Narita}, {Paegert}, {Palle}, {Pepe},
  {Pepper}, {Quirrenbach}, {Rinehart}, {Sasselov}, {Sato}, {Seager},
  {Sozzetti}, {Stassun}, {Sullivan}, {Szentgyorgyi}, {Torres}, {Udry}, \&
  {Villasenor}}]{2014SPIE.9143E..20R}
{Ricker}, G.~R., {Winn}, J.~N., {Vanderspek}, R., {et~al.} 2014, in Society of
  Photo-Optical Instrumentation Engineers (SPIE) Conference Series, Vol. 9143,
  Space Telescopes and Instrumentation 2014: Optical, Infrared, and Millimeter
  Wave, ed. J.~{Oschmann}, Jacobus~M., M.~{Clampin}, G.~G. {Fazio}, \& H.~A.
  {MacEwen}, 914320, \dodoi{10.1117/12.2063489}

\bibitem[{{Rieger}(2004)}]{2004ApJ...615L...5R}
{Rieger}, F.~M. 2004, \apjl, 615, L5, \dodoi{10.1086/426018}

\bibitem[{{Romero} {et~al.}(2000){Romero}, {Chajet}, {Abraham}, \&
  {Fan}}]{2000A&A...360...57R}
{Romero}, G.~E., {Chajet}, L., {Abraham}, Z., \& {Fan}, J.~H. 2000, \aap, 360,
  57

\bibitem[{{Roy} {et~al.}(2022{\natexlab{a}}){Roy}, {Sarkar}, {Chatterjee},
  {Gupta}, {Chitnis}, \& {Wiita}}]{2022MNRAS.510.3641R}
{Roy}, A., {Sarkar}, A., {Chatterjee}, A., {et~al.} 2022{\natexlab{a}}, \mnras,
  510, 3641, \dodoi{10.1093/mnras/stab3701}

\bibitem[{{Roy} {et~al.}(2022{\natexlab{b}}){Roy}, {Chitnis}, {Gupta}, {Wiita},
  {Romero}, {Cellone}, {Chatterjee}, {Combi}, {Raiteri}, {Sarkar}, \&
  {Villata}}]{2022MNRAS.513.5238R}
{Roy}, A., {Chitnis}, V.~R., {Gupta}, A.~C., {et~al.} 2022{\natexlab{b}},
  \mnras, 513, 5238, \dodoi{10.1093/mnras/stac1287}

\bibitem[{{Sandrinelli} {et~al.}(2016){Sandrinelli}, {Covino}, {Dotti}, \&
  {Treves}}]{2016AJ....151...54S}
{Sandrinelli}, A., {Covino}, S., {Dotti}, M., \& {Treves}, A. 2016, \aj, 151,
  54, \dodoi{10.3847/0004-6256/151/3/54}

\bibitem[{{Sandrinelli} {et~al.}(2018){Sandrinelli}, {Covino}, {Treves},
  {Holgado}, {Sesana}, {Lindfors}, \& {Ramazani}}]{2018A&A...615A.118S}
{Sandrinelli}, A., {Covino}, S., {Treves}, A., {et~al.} 2018, \aap, 615, A118,
  \dodoi{10.1051/0004-6361/201732550}

\bibitem[{{Sarkar} {et~al.}(2021){Sarkar}, {Gupta}, {Chitnis}, \&
  {Wiita}}]{2021MNRAS.501...50S}
{Sarkar}, A., {Gupta}, A.~C., {Chitnis}, V.~R., \& {Wiita}, P.~J. 2021, \mnras,
  501, 50, \dodoi{10.1093/mnras/staa3211}

\bibitem[{{Sarkar} {et~al.}(2020){Sarkar}, {Kushwaha}, {Gupta}, {Chitnis}, \&
  {Wiita}}]{2020A&A...642A.129S}
{Sarkar}, A., {Kushwaha}, P., {Gupta}, A.~C., {Chitnis}, V.~R., \& {Wiita},
  P.~J. 2020, \aap, 642, A129, \dodoi{10.1051/0004-6361/202038052}

\bibitem[{{Scargle}(1982)}]{1982ApJ...263..835S}
{Scargle}, J.~D. 1982, \apj, 263, 835, \dodoi{10.1086/160554}

\bibitem[{{Stella} \& {Vietri}(1998)}]{1998ApJ...492L..59S}
{Stella}, L., \& {Vietri}, M. 1998, \apjl, 492, L59, \dodoi{10.1086/311075}

\bibitem[{{Stickel} {et~al.}(1991){Stickel}, {Padovani}, {Urry}, {Fried}, \&
  {Kuehr}}]{1991ApJ...374..431S}
{Stickel}, M., {Padovani}, P., {Urry}, C.~M., {Fried}, J.~W., \& {Kuehr}, H.
  1991, \apj, 374, 431, \dodoi{10.1086/170133}

\bibitem[{{Stocke} {et~al.}(1991){Stocke}, {Morris}, {Gioia}, {Maccacaro},
  {Schild}, {Wolter}, {Fleming}, \& {Henry}}]{1991ApJS...76..813S}
{Stocke}, J.~T., {Morris}, S.~L., {Gioia}, I.~M., {et~al.} 1991, \apjs, 76,
  813, \dodoi{10.1086/191582}

\bibitem[{{Tanaka} {et~al.}(2016){Tanaka}, {Becerra Gonzalez}, {Itoh}, {Finke},
  {Inoue}, {Ojha}, {Carpenter}, {Lindfors}, {Krau{\ss}}, {Desiante}, {Shiki},
  {Fukazawa}, {Longo}, {McEnery}, {Buson}, {Nilsson}, {Fallah Ramazani},
  {Reinthal}, {Takalo}, {Pursimo}, \& {Boschin}}]{2016PASJ...68...51T}
{Tanaka}, Y.~T., {Becerra Gonzalez}, J., {Itoh}, R., {et~al.} 2016, \pasj, 68,
  51, \dodoi{10.1093/pasj/psw049}

\bibitem[{{Templeton}(2004)}]{2004JAVSO..32...41T}
{Templeton}, M. 2004, \jaavso, 32, 41

\bibitem[{{Tripathi} {et~al.}(2021){Tripathi}, {Gupta}, {Aller}, {Wiita},
  {Bambi}, {Aller}, \& {Gu}}]{2021MNRAS.501.5997T}
{Tripathi}, A., {Gupta}, A.~C., {Aller}, M.~F., {et~al.} 2021, \mnras, 501,
  5997, \dodoi{10.1093/mnras/stab058}

\bibitem[{Twicken {et~al.}(2020)Twicken, Caldwell, Jenkins, Vanderspek,
  Tenenbaum, Smith, Wohler, Rose, Ting, Vanderspek, {et~al.}}]{twicken2020tess}
Twicken, J.~D., Caldwell, D.~A., Jenkins, J.~M., {et~al.} 2020, \

\bibitem[{{Urry} \& {Padovani}(1995)}]{1995PASP..107..803U}
{Urry}, C.~M., \& {Padovani}, P. 1995, \pasp, 107, 803, \dodoi{10.1086/133630}

\bibitem[{{Valtonen} {et~al.}(2008){Valtonen}, {Lehto}, {Nilsson}, {Heidt},
  {Takalo}, {Sillanp{\"a}{\"a}}, {Villforth}, {Kidger}, {Poyner}, {Pursimo},
  {Zola}, {Wu}, {Zhou}, {Sadakane}, {Drozdz}, {Koziel}, {Marchev}, {Ogloza},
  {Porowski}, {Siwak}, {Stachowski}, {Winiarski}, {Hentunen}, {Nissinen},
  {Liakos}, \& {Dogru}}]{2008Natur.452..851V}
{Valtonen}, M.~J., {Lehto}, H.~J., {Nilsson}, K., {et~al.} 2008, \nat, 452,
  851, \dodoi{10.1038/nature06896}

\bibitem[{{Villforth} {et~al.}(2010){Villforth}, {Nilsson}, {Heidt}, {Takalo},
  {Pursimo}, {Berdyugin}, {Lindfors}, {Pasanen}, {Winiarski}, {Drozdz},
  {Ogloza}, {Kurpinska-Winiarska}, {Siwak}, {Koziel-Wierzbowska}, {Porowski},
  {Kuzmicz}, {Krzesinski}, {Kundera}, {Wu}, {Zhou}, {Efimov}, {Sadakane},
  {Kamada}, {Ohlert}, {Hentunen}, {Nissinen}, {Dietrich}, {Assef}, {Atlee},
  {Bird}, {Depoy}, {Eastman}, {Peeples}, {Prieto}, {Watson}, {Yee}, {Liakos},
  {Niarchos}, {Gazeas}, {Dogru}, {Donmez}, {Marchev}, {Coggins-Hill},
  {Mattingly}, {Keel}, {Haque}, {Aungwerojwit}, \&
  {Bergvall}}]{2010MNRAS.402.2087V}
{Villforth}, C., {Nilsson}, K., {Heidt}, J., {et~al.} 2010, \mnras, 402, 2087,
  \dodoi{10.1111/j.1365-2966.2009.16133.x}

\bibitem[{{Wagner} \& {Witzel}(1995)}]{1995ARA&A..33..163W}
{Wagner}, S.~J., \& {Witzel}, A. 1995, \araa, 33, 163,
  \dodoi{10.1146/annurev.aa.33.090195.001115}

\bibitem[{{Wagner} {et~al.}(1993){Wagner}, {Witzel}, {Krichbaum}, {Wegner},
  {Quirrenbach}, {Anton}, {Erkens}, {Khanna}, \&
  {Zensus}}]{1993A&A...271..344W}
{Wagner}, S.~J., {Witzel}, A., {Krichbaum}, T.~P., {et~al.} 1993, \aap, 271,
  344

\bibitem[{{Zhang} {et~al.}(2017{\natexlab{a}}){Zhang}, {Zhang}, {Yan}, {Fan},
  \& {Liu}}]{2017ApJ...849....9Z}
{Zhang}, P., {Zhang}, P.-f., {Yan}, J.-z., {Fan}, Y.-z., \& {Liu}, Q.-z.
  2017{\natexlab{a}}, \apj, 849, 9, \dodoi{10.3847/1538-4357/aa8d6e}

\bibitem[{{Zhang} {et~al.}(2017{\natexlab{b}}){Zhang}, {Yan}, {Zhou}, {Fan},
  {Wang}, \& {Zhang}}]{2017ApJ...845...82Z}
{Zhang}, P.-F., {Yan}, D.-H., {Zhou}, J.-N., {et~al.} 2017{\natexlab{b}}, \apj,
  845, 82, \dodoi{10.3847/1538-4357/aa7ecd}

\bibitem[{{Zhang} {et~al.}(2018){Zhang}, {Zhang}, {Liao}, {Yan}, {Fan}, \&
  {Liu}}]{2018ApJ...853..193Z}
{Zhang}, P.-f., {Zhang}, P., {Liao}, N.-h., {et~al.} 2018, \apj, 853, 193,
  \dodoi{10.3847/1538-4357/aaa29a}

\bibitem[{{Zhang} \& {Bao}(1991)}]{1991A&A...246...21Z}
{Zhang}, X.~H., \& {Bao}, G. 1991, \aap, 246, 21

\end{thebibliography}
\section*{ Supplimentary materials}
\subsection*{Sector-wise binned lightcurves}

\newpage

\subsection*{Segment-wise GLSP and wavelet transform in each sector}
\newpage
Fig. 6 continued...

\newpage
Fig. 6 continued...
\newpage
Fig. 6 continued..
\newpage
Fig. 6 continued..
\newpage
Fig. 6 continued..

\bibliographystyle{aasjournal}
\end{document}